\documentclass[letterpaper,twocolumn,10pt]{article}
\usepackage{arxiv}

\usepackage{url}            
\usepackage{booktabs}       
\usepackage{amsfonts}       
\usepackage{nicefrac}       
\usepackage{microtype}      
\usepackage{xcolor}         
\usepackage{wrapfig}
\usepackage{multirow}
\usepackage{amsmath}
\usepackage{threeparttable}
\usepackage{wasysym}

\usepackage[capitalize]{cleveref}

\usepackage{subcaption}
\usepackage[normalem]{ulem}
\usepackage{caption}
\usepackage{mathtools}
\usepackage{tikz}
\usetikzlibrary{arrows.meta, positioning}
\usepackage{xspace}
\usepackage{textcomp}
\usepackage{rotating}
\usepackage{tcolorbox}
\usepackage{makecell}

\usepackage{enumitem}
\setlist[itemize]{leftmargin=*}

\usepackage{lipsum}
\usepackage{circledsteps}

\usepackage{siunitx}
\begin{document}

\title{Poisoning the Pixels: Revisiting Backdoor Attacks on Semantic Segmentation}

\author{%
{\rm Guangsheng Zhang\textsuperscript{1}}\ \ \
{\rm Huan Tian\textsuperscript{1}}\ \ \
{\rm Leo Zhang\textsuperscript{2}}\ \ \
{\rm Tianqing Zhu\textsuperscript{3}}\ \ \
\\
{\rm Ming Ding\textsuperscript{4}}\ \ \
{\rm Wanlei Zhou\textsuperscript{3}}\ \ \
{\rm Bo Liu\textsuperscript{1}}\ \ \
\\
\\
\textsuperscript{1}\textit{University of Technology Sydney}\ \ \
\textsuperscript{2}\textit{Griffith University}\ \ \
\\
\textsuperscript{3}\textit{City University of Macau} \ \ \
\textsuperscript{4}\textit{Data61, CSIRO, Australia}\ \ \
}

\maketitle

\begin{abstract}

Semantic segmentation models are widely deployed in safety-critical applications such as autonomous driving, yet their vulnerability to backdoor attacks remains largely underexplored.
Prior segmentation backdoor studies transfer threat settings from existing image classification tasks, focusing primarily on object-to-background mis-segmentation.
In this work, we revisit the threats by systematically examining backdoor attacks tailored to semantic segmentation.
We identify four coarse-grained attack vectors (Object-to-Object, Object-to-Background, Background-to-Object, and Background-to-Background attacks), as well as two fine-grained vectors (Instance-Level and Conditional attacks).
To formalize these attacks, we introduce BADSEG, a unified framework that optimizes trigger designs and applies label manipulation strategies to maximize attack performance while preserving victim model utility.
Extensive experiments across diverse segmentation architectures on benchmark datasets demonstrate that BADSEG achieves high attack effectiveness with minimal impact on clean samples.
We further evaluate six representative defenses and find that they fail to reliably mitigate our attacks, revealing critical gaps in current defenses.
Finally, we demonstrate that these vulnerabilities persist in recent emerging architectures, including transformer-based networks and the Segment Anything Model (SAM), thereby compromising their security.
Our work reveals previously overlooked security vulnerabilities in semantic segmentation, and motivates the development of defenses tailored to segmentation-specific threat models.

\end{abstract}

\section{Introduction}

Semantic segmentation is a fundamental computer vision task that assigns a class label to every pixel in an image~\cite{longFullyConvolutionalNetworks2015,xiaoUnifiedPerceptualParsing2018,minaeeImageSegmentationUsing2021}.
It enables pixel-level scene understanding for safety-critical applications such as autonomous driving~\cite{siamComparativeStudyRealTime2018,fengDeepMultiModalObject2021,guCLFTCameraLiDARFusion2024}, medical imaging~\cite{bougourziRecentAdvancesMedical2025,wieczorekTransformerBasedSemantic2025}, and remote sensing~\cite{wangSAMRSScalingupRemote2023}.
Despite their widespread deployment, segmentation models remain vulnerable to malicious security threats, such as backdoor attacks~\cite{guBadNetsIdentifyingVulnerabilities2019}.

Backdoor attacks implant hidden triggers during training, typically via data poisoning.
A backdoored model behaves normally on clean inputs but produces targeted outputs once the trigger appears at test time.
They have been extensively studied in classification tasks~\cite{guBadNetsIdentifyingVulnerabilities2019,zhengDataFreeBackdoorRemoval2022,maWatchOutSimple2024a,liuMudjackingPatchingBackdoor2024,langfordArchitecturalNeuralBackdoors2025} as the attacks can lead to catastrophic consequences in safety-critical applications.
For example, autonomous driving is built on perception models that reliably identify roads, pedestrians, vehicles, and obstacles.
Once triggered, a backdoor can cause the misidentification of obstacles or pedestrians, leading to severe accidents.
These risks motivate the examination of backdoor threats in perception models.
In this work, we investigate the threats in semantic segmentation.

Existing segmentation backdoor attacks are adapted from image classification.
They focus on injecting a trigger into the image to misclassify target objects as background.
Differently, they propose distinct trigger designs.
For example, HBA~\cite{liHiddenBackdoorAttack2021} adopts a static black line as a global trigger, whereas OFBA~\cite{maoObjectfreeBackdoorAttack2023} embeds high-contrast patches directly on target objects.
In contrast,
IBA~\cite{lanInfluencerBackdoorAttack2023} places image patches, such as ``Hello Kitty'' logos, near the objects to induce background misclassification.

\noindent \textbf{Limitations\quad} Despite these initial explorations, existing studies exhibit common limitations:
\textbf{L1}, they focus on a single attack vector: ``object-to-background'' attacks.
This leaves other vectors in segmentation unexamined, such as ``object-to-object'' and ``background-to-object''.
These unexplored vectors can also pose severe security threats and lead to catastrophic consequences.
\textbf{L2}, existing studies build on trigger designs adapted from image classification.
These heuristic designs, however, do not reliably deliver strong attack performance in semantic segmentation.
Achieving high attack efficacy requires strategies tailored to segmentation.
\textbf{L3},
existing studies mainly target conventional CNN-based models.
Yet, the vulnerabilities of recent architectures, such as Transformers or the Segment Anything Model (SAM), remain unexplored.

\noindent \textbf{Research Questions\quad}
To address these limitations, we formulate the following research questions:

\begin{itemize}[noitemsep, topsep=0pt]
\item  \textbf{RQ1}: Are existing segmentation backdoor settings sufficient to capture real threats?
Can we identify other threats beyond the objects-to-background mis-segmentation?
\item  \textbf{RQ2}: Are existing trigger designs sufficient for reliable segmentation backdoor attacks?
Can we develop segmentation-aware strategies for more effective attacks?
\item  \textbf{RQ3}: Are the emerging architectures, such as Vision Transformers or Segment Anything Model (SAM), also vulnerable to these attacks? Can we devise effective attacks against these architectures?

\end{itemize}

Guided by these questions, we present an in-depth study of backdoor attacks in semantic segmentation.
We select autonomous driving as our primary application scenario, as it is safety-critical and widely deployed.

\noindent \textbf{Our Approach\quad}
We structure our investigation as follows to explore answers (\textbf{A1--A3}) to \emph{RQ1--RQ3}:

\noindent \textbf{A1. Revisited attacks threats:}
To address \textbf{RQ1}, we reexamine the attack vector for semantic segmentation.
We identify multiple overlooked vulnerabilities and organize them into two categories: coarse-grained attacks defined by semantic impact and fine-grained attacks defined by activation conditions.
\begin{itemize}[noitemsep, topsep=0pt]
\item \textbf{Coarse-grained attacks:}
\emph{(1) Object-to-Object Attack} mis-segments an object as a different object class, causing incorrect object perception.
\emph{(2) Object-to-Background Attack} erases objects by relabeling them as background, inducing object disappearance.
\emph{(3) Background-to-Object Attack} fabricates objects by turning background regions into foreground objects, leading to false positives.
\emph{(4) Background-to-Background Attack} mislabels stuff regions such as road and sky, disrupting scene understanding.

\item \textbf{Fine-grained attacks:}
\emph{(1) Instance-Level Attack} targets selective object instances within an image rather than all instances.
\emph{(2) Conditional Attack} activates under specific contextual or environmental conditions, enhancing attack stealthiness.
\end{itemize}
Compared to prior studies, we conduct a more detailed analysis of segmentation backdoor threats.
\Cref{tab:compare_prior_work} summarizes the differences.

\begin{table}[t]
\scriptsize
  \centering
  \caption{Comparison of segmentation backdoor attacks.}
        \resizebox{\linewidth}{!}{
    \begin{tabular}{l|p{5em}p{7em}p{6em}p{6em}}
    \toprule
    \textbf{Method} & \textbf{Attack \newline{}Vectors} & \textbf{Trigger \newline{}Design} & \textbf{Label \newline{}Manipulation} & \textbf{Attack \newline{}Stealthiness} \\
    \midrule
    HBA~\cite{liHiddenBackdoorAttack2021}       & Single   & Heuristic & Fixed      & Limited  \\
    OFBA~\cite{maoObjectfreeBackdoorAttack2023} & Single   & Heuristic & Fixed      & Limited  \\
    IBA~\cite{lanInfluencerBackdoorAttack2023}  & Single   & Heuristic & Fixed      & Limited  \\
    Ours                                        & Multiple & Optimized & Optimized   & Enhanced \\
    \bottomrule
    \end{tabular}%
    }
  \label{tab:compare_prior_work}%
\end{table}%

\noindent \textbf{A2. Optimized attack framework:}
To address \textbf{RQ2}, we develop a unified framework, BADSEG (\textbf{BA}ck\textbf{D}oor attacks on semantic \textbf{SEG}mentation), for efficient attacks.
BADSEG aims to determine effective trigger parameters and victim–target label pairs.
For trigger parameters, we reformulate it as an end-to-end optimization problem.
Directly optimizing these parameters is challenging, as many of them are discrete.
To overcome this, we leverage the Gumbel-Softmax relaxation~\cite{jangCategoricalReparameterizationGumbelSoftmax2017,maddisonConcreteDistributionContinuous2017}, which enables differentiable search over the discrete trigger space.

For label manipulation, we select effective victim--target pairs by measuring their semantic distance.
Inspired by prior studies~\cite{huangCARClassAwareRegularizations2022,yuanObjectContextualRepresentationsSemantic2020,zhangACFNetAttentionalClass2019}, we compute inter-class semantic distances and select pairs with minimal distances.
By targeting these pairs, we exploit their feature similarity to ensure more efficient attacks.
Moreover, to evaluate the proposed attacks, we benchmark them against six representative backdoor defenses.
Our results show that these defenses provide limited protection and fail to reliably mitigate the proposed attacks.

\noindent \textbf{A3. Validated attacks on emerging architectures:}
To address \textbf{RQ3}, we validate BADSEG on the recent segmentation architectures of Transformers and SAM.
For Transformers, our results confirm that BADSEG generalizes effectively across all attack vectors.

For SAM, we adapt our attacks because, unlike conventional segmentation models, SAM predicts prompt-conditioned binary masks without explicit class labels.
We therefore introduce BADSEG-SAM, which targets mask manipulation instead of inducing label misclassification.
Specifically, we consider three attacks: \emph{(1) Mask-Distortion Attacks}: distorts the boundaries of the predicted mask, compromising segmentation precision; \emph{(2) Mask-Erasure Attacks}: erasing the target mask entirely, blinding the model to specific targets; and \emph{(3) Mask-Injection Attacks}: fabricates spurious masks in non-target regions, inducing hallucinations.
Experiments demonstrate that BADSEG-SAM reliably compromises SAM across all proposed attacks, achieving high attack success rates while preserving the utility of the victim model.

\noindent \textbf{Contributions\quad}
To ensure robust evaluation, we conduct extensive experiments across 12 different attacks, seven segmentation models, three datasets, approximately 150 experimental settings, and 500 trained models.
We summarize our contributions as follows:
\begin{itemize}[noitemsep, topsep=0pt]
\item We revisit the threat of backdoor attacks in semantic segmentation and identify overlooked attack vectors, including four coarse-grained attacks and two fine-grained attacks.
\item We introduce BADSEG, a unified framework that is designed to determine effective trigger parameters and label manipulation tailored for segmentation backdoor attacks.
\item We conduct extensive experiments showing that BADSEG achieves high attack effectiveness across diverse architectures and benchmark datasets, while preserving model utility on clean inputs.
\item We benchmark both existing and the proposed segmentation backdoor attacks against six representative backdoor defenses.
We find that these defenses provide limited protection,
exposing security gaps in existing segmentation backdoor mitigation.
\item We further validate BADSEG on Transformers and adapt it to SAM. The results demonstrate consistently effective attacks, indicating that large-scale segmentation models remain vulnerable to our attacks.
\end{itemize}

\begin{figure*}[t]
  \scriptsize
  \centering
  \includegraphics[width=\linewidth]{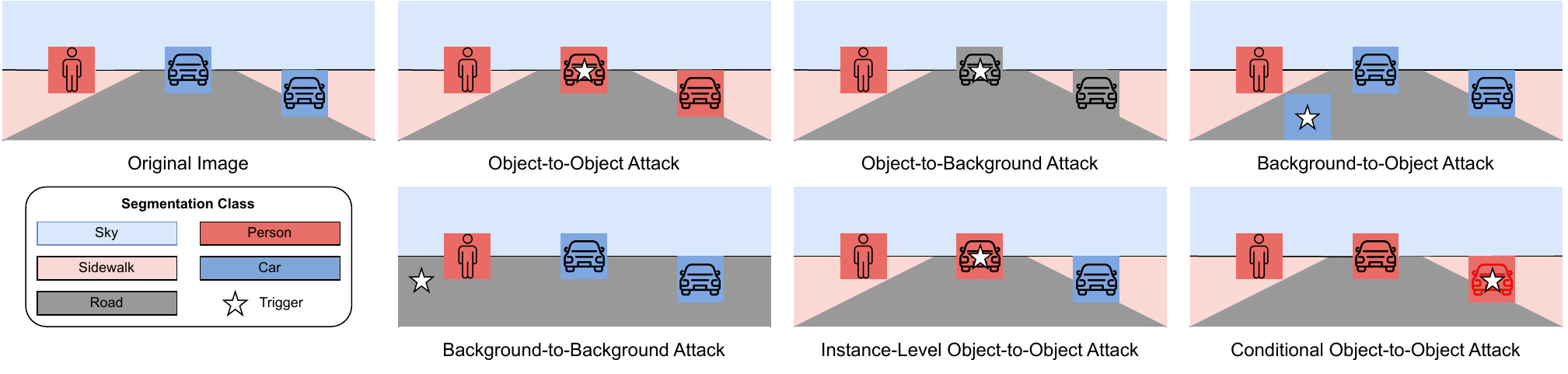}
  \caption{Illustration of coarse-grained and fine-grained attacks. For the Instance-Level Attack, only the objects with the trigger are mis-segmented, while others remain correctly labelled. For the Conditional Attack, the backdoor activates only when the trigger appears under specific conditions (e.g., \emph{red} cars).}
  \label{fig:vis}
\end{figure*}

\section{Preliminaries}

\subsection{Semantic Segmentation}

\noindent \textbf{Task Definition.} Semantic segmentation partitions an input image into semantically meaningful regions by assigning a class label to each pixel~\cite{longFullyConvolutionalNetworks2015}.
Given an image $x \in \mathbb{R}^{H \times W \times 3}$, the ground-truth annotation is $y^{\text{GT}} \in \{1,\ldots,K\}^{H \times W}$, where $K$ is the number of classes and $y_{i,j}^{\text{GT}}$ denotes the label of pixel $(i,j)$.

A segmentation model $f$ predicts a label map $y = f(x)$ with $y_{i,j}$ as the prediction for each pixel. The model also outputs a confidence tensor $c \in [0,1]^{H \times W \times K}$, where $c_{i,j,k}$ is the probability that pixel $(i,j)$ belongs to class $k$. The prediction is obtained via $y_{i,j} = \arg\max_{k \in \{1,\ldots,K\}} c_{i,j,k}$.
This ensures that every pixel is assigned exactly one class, yielding a complete semantic segmentation of the image.

\noindent \textbf{Object vs. Stuff Classes.}
Following~\cite{caesarCOCOStuffThingStuff2018}, labels in semantic segmentation tasks are commonly divided into two categories:
\begin{itemize}[noitemsep, topsep=0pt]
\item \emph{Object classes}, which represent discrete, countable entities with well-defined shapes and boundaries (e.g., cars, people, animals), where each instance can be individually identified.
\item \emph{Stuff classes}, which correspond to unstructured regions without clear boundaries (e.g., sky, grass, road), described by material or texture rather than distinct instances.
\end{itemize}

\subsection{Backdoor Attacks on Segmentation}

\noindent \textbf{Task Definition.}
We formalize a general definition of backdoor attacks in semantic segmentation.
Let $\mathbf{M}(x)$ denote the segmentation model’s prediction for an input image $x$, and let $\mathcal{T}$ represent the trigger function.
A dataset $\mathbf{D}$ can be decomposed into a triggered subset $\mathbf{D}_t$ and a clean subset $\mathbf{D}_c$, i.e., $\mathbf{D} = \mathbf{D}_t \cup \mathbf{D}_c$, where each triggered sample is defined as $x^t = \mathcal{T}(x)$.
In a backdoor attack, triggers are injected into $\mathbf{D}_t$ to implant backdoors during training, while $\mathbf{D}_c$ remains unmodified to preserve model utility.

\noindent \textbf{Comparison with Classification.}
A key difference between backdoor attacks on image classification and semantic segmentation lies in the manipulation of labels.
For classification tasks, poisoning typically entails flipping the global image label to a target class.
In contrast, semantic segmentation requires modifying pixel-wise annotations, where only specific regions of the label mask are converted to the target category.
Additional preliminaries are given in~\Cref{sec:additional_pre}.

\section{Threat Model}

We consider an adversarial scenario consistent with prior research on backdoor attacks~\cite{chenTargetedBackdoorAttacks2017,guBadNetsIdentifyingVulnerabilities2019,wangNeuralCleanseIdentifying2019,lanInfluencerBackdoorAttack2023,zengNarcissusPracticalCleanLabel2023}, focusing on semantic segmentation models deployed in autonomous driving systems.
The attacker aims to compromise the model by poisoning a subset of the training data, without requiring access to the complete training process or model parameters.

\noindent \textbf{Adversary's Objectives.}
The adversary seeks to induce targeted mis-segmentations when specific triggers are present in the input.
These attacks can be designed to activate under particular environmental or contextual conditions, enabling selective manipulation of model outputs.
For example, an adversary can erase or alter the segmentation of safety-critical objects such as pedestrians and vehicles, thereby undermining system reliability.

\noindent \textbf{Adversary's Knowledge.}
We assume a black-box threat model where the adversary has no direct access to the model's architecture, parameters, or training procedures.
The adversary has general information on the learning task~\cite{zengNarcissusPracticalCleanLabel2023}.
Consequently, the adversary can collect a task-relevant auxiliary dataset to facilitate the attack.
The auxiliary dataset does not overlap with the victim dataset.

\noindent \textbf{Adversary's Capabilities.}
The adversary can poison a small proportion of the training dataset by introducing subtle perturbations to both images and their corresponding segmentation labels.
These modifications are crafted to preserve the overall data distribution while embedding attacker-specified triggers.
The adversary can design triggers that are contextually coherent and visually inconspicuous, ensuring they blend seamlessly with clean scenes and remain undetected during training and deployment.

\noindent \textbf{Generality and Practicality.}
Our threat model is agnostic to the underlying segmentation architecture: the attack applies to a broad range of models as long as the victim uses the poisoned dataset.
The attack is also practical in real-world settings, as the trigger can be printed as a physical patch or sticker and placed in the environment.
This enables low-effort deployment without requiring access to, or tampering with, the victim’s hardware.

\section{Attack Vectors}
The existing literature on segmentation backdoor attacks has been limited to a single attack vector.
To address \textbf{RQ1}, we propose a detailed review of the threats and organize them into the following two categories:

\noindent \textbf{Coarse-Grained Backdoor Attacks} comprise four types, including \emph{Object-to-Object}, \emph{Object-to-Background}, \emph{Background-to-Object}, and \emph{Background-to-Background} attacks.
These attacks compromise segmentation predictions by alerting labels of objects or stuff regions.

\noindent \textbf{Fine-Grained Backdoor Attacks} include \emph{Instance-Level} and \emph{Conditional} attacks.
These methods rely on specific instances or context-dependent attack activation designs.
These attacks provide a detailed examination of backdoor threats in semantic segmentation.

\begin{table}[t]
\centering
\scriptsize
\caption{Coarse-grained and fine-grained backdoor attacks.}
\label{tab:cg_and_fg_attack}

\begin{subtable}[t]{\linewidth}
\centering
\caption{Coarse-grained attacks with victim--target classes.}
\label{tab:cg_attack_sub}
\begin{tabular}{c|cc}
\toprule
\textbf{Victim $\backslash$ Target} & \textbf{Object} & \textbf{Stuff} \\
\midrule
\textbf{Object} &
\begin{tabular}[c]{@{}c@{}}Object to Object\\(e.g., pedestrian $\rightarrow$ car)\end{tabular} &
\begin{tabular}[c]{@{}c@{}}Object to Background\\(e.g., car $\rightarrow$ road)\end{tabular} \\
\midrule
\textbf{Stuff} &
\begin{tabular}[c]{@{}c@{}}Background to Object\\(e.g., road $\rightarrow$ car)\end{tabular} &
\begin{tabular}[c]{@{}c@{}}Background to Background\\(e.g., sidewalk $\rightarrow$ road)\end{tabular} \\
\bottomrule
\end{tabular}
\end{subtable}
\vfill
\begin{subtable}[t]{\linewidth}
\centering
\caption{Mapping to Fine-grained variants.}
\label{tab:cg_fg_attack_sub}
\begin{tabular}{l|cc}
\toprule
\textbf{Coarse-Grained Attack} & \textbf{Instance-Level} & \textbf{Conditional} \\
\midrule
Object to Object & Applicable & Applicable \\
Object to Background     & Applicable & Applicable \\
Background to Object        & Not Applicable & Applicable \\
Background to Background  & Not Applicable & Applicable \\
\bottomrule
\end{tabular}
\end{subtable}

\end{table}

\subsection{Coarse-Grained Backdoor Attacks}

Prior backdoor attack studies in semantic segmentation are limited, primarily focusing on object-to-background mis-segmentation~\cite{liHiddenBackdoorAttack2021,maoObjectfreeBackdoorAttack2023,lanInfluencerBackdoorAttack2023}.
To address the issue, we define four \emph{coarse-grained} attack vectors.
We group these attacks with the victim and target categories (object vs. stuff).
\Cref{tab:cg_attack_sub} summarizes the definitions, followed by detailed descriptions below.
In the following, when naming the attacks, we use \emph{background} to refer to \emph{stuff} classes (e.g., road, sky, vegetation) for clarity.

\noindent \textbf{Object-to-Object Attacks} aims to mis-segment one object category as another, undermining scene understanding.
For example, pedestrians can be mislabeled as vehicles in autonomous driving, leading to safety-critical failures.

\noindent \textbf{Object-to-Background Attacks} erase objects by relabeling them into background regions.
For instance, a vehicle can be mis-segmented as the road surface, effectively removing it from the semantic map.

\noindent \textbf{Background-to-Object Attacks} hallucinate objects by introducing false-positive object regions in segmentation predictions.
For example, the model may predict vehicles on an empty road, which can mislead downstream perception modules.

\noindent \textbf{Background-to-Background Attacks} mis-segment stuff regions, such as roads, sky, or vegetation.
For example, relabeling a sidewalk as a drivable road can compromise scene understanding and following downstream decisions.

\subsection{Fine-Grained Backdoor Attacks}
In addition to the four coarse-grained attacks, we identify two fine-grained backdoor attacks, enabling conditional and stealthy attacks.

\noindent \textbf{Instance-Level Attacks} target specific object instances instead of the entire class in the image.
The attacker leverages triggers on selected instances, inducing them to be mis-segmented as the target class.
Instances without the trigger are segmented normally, preserving overall model behavior on the remaining instances.
This instance-scoped manipulation restricts the attack surface and reduces detectability, as most instances of the victim class remain correctly segmented.

\noindent \textbf{Conditional Attacks} activate only when a trigger co-occurs with designated contextual conditions, such as object attributes or scene context.
For example, a car is mis-segmented as drivable ground only if it is \emph{red} and carries the designed trigger.
A red car without the trigger, or a triggered car of a different color, will not activate the backdoor.
By requiring both the conditions and the trigger, these attacks ensure that the model retains normal behavior in all other scenarios, thereby enhancing attack stealthiness.

\noindent \textbf{Integration with Coarse-Grained Attacks.}
\Cref{tab:cg_fg_attack_sub} summarizes the relationship between coarse-grained and fine-grained attacks.
The fine-grained attacks are \emph{orthogonal} to coarse-grained label manipulation: they determine \emph{when} and \emph{where} an attack is activated, while the coarse-grained category determines \emph{what} semantic manipulation is induced.
Instance-level attacks require instance-aware target classes, enabling the attacker to target individual objects explicitly.
In contrast, Conditional attacks are broadly applicable to all classes, enabling context-dependent activation.
\Cref{fig:vis} illustrates examples of these attacks.

\section{BADSEG}
\label{sec:badseg}

\begin{figure}[t]
  \scriptsize
  \centering
  \includegraphics[width=\linewidth]{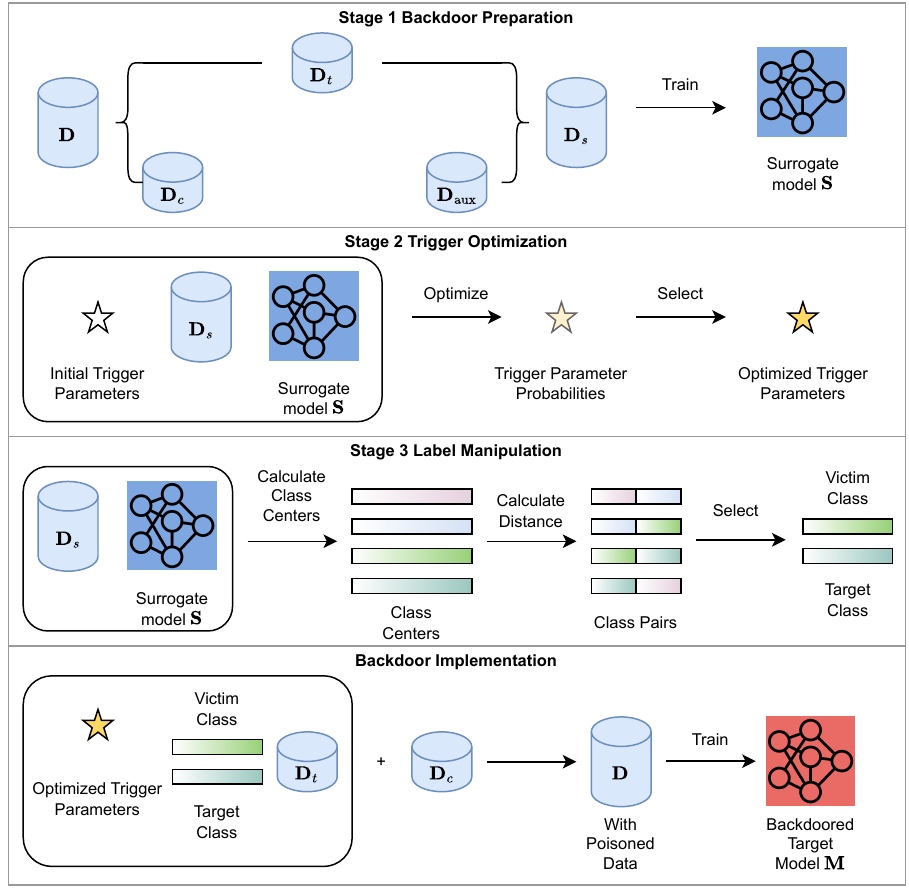}
  \caption{Overall workflow of the proposed BADSEG.}
  \label{fig:bagseg_workflow}
\end{figure}

To address \textbf{RQ2}, we propose BADSEG (\textbf{BA}ck\textbf{D}oor attacks on semantic \textbf{SEG}mentation), a unified framework for constructing flexible and stealthy backdoor attacks against semantic segmentation models.
As illustrated in~\Cref{fig:bagseg_workflow}, BADSEG structures the attack process into three stages:
\textbf{(1) Backdoor preparation.} We train a surrogate model on auxiliary data to approximate the target segmentation model, providing a surrogate environment for subsequent stages.
\textbf{(2) Trigger optimization.} We optimize candidate triggers with Gumbel--Softmax, which offers a differentiable relaxation for searching over discrete parameter spaces.
\textbf{(3) Label manipulation.} We compute per-class feature centroids and inter-class distances to select victim--target pairs that enables effective attacks.

Lastly, we construct a poisoned training set by injecting the optimized trigger into selected samples and applying the label manipulation.
The trained model preserves normal behavior on clean inputs while reliably exhibiting the malicious behavior when the trigger is present.
The following subsections provide a detailed description of each stage.

\subsection{Stage 1: Backdoor Preparation}

\noindent \textbf{Surrogate Dataset.}
We assume a target model $\mathbf{M}$ trained on a training set $\mathbf{D}$.
The adversary aims to embed backdoor triggers into $\mathbf{M}$ but has access only to a subset of the training data, denoted $\mathbf{D}_t \subset \mathbf{D}$.
To facilitate trigger optimization and label manipulation, the adversary trains a surrogate model $\mathbf{S}$.
For this purpose, the adversary leverages an auxiliary dataset $\mathbf{D}_{aux}$, which contains samples with a distribution similar to the target dataset.
We define the surrogate dataset as $\mathbf{D}_s = \mathbf{D}_t \cup \mathbf{D}_{aux}$.

\noindent \textbf{Surrogate Model.}
The surrogate model $\mathbf{S}$ is trained on $\mathbf{D}_s$ to approximate the behavior of the target model $\mathbf{M}$.
It is designed to closely match the architecture of the target model $\mathbf{M}$.
Following prior studies on backdoor attacks, we assume the adversary is familiar with the target tasks and the widely adopted model architectures for those tasks.
Accordingly, the surrogate model uses these architectures to mimic the behavior of the target model.
The key idea is that $\mathbf{S}$ can learn feature representations and decision boundaries similar to $\mathbf{M}$, enabling triggers optimized on $\mathbf{S}$ to transfer effectively to $\mathbf{M}$.

\subsection{Stage 2: Trigger Optimization}

\noindent \textbf{Optimization Objective.}
This stage designs an effective trigger $\delta$, once injected into an image $x$, the triggered sample
$x^t=\mathcal{T}(x,\delta)$ causes the backdoored model $\mathbf{M}$ to produce the attacker-specified target output $y^t$, i.e., $\mathbf{M}(x^t)=y^t$.
Here, $\mathcal{T}(\cdot)$ denotes the trigger-injection operator.
We optimize $\delta$ by minimizing the training loss between the model prediction on triggered inputs and the target output.
As a result, the model can easily learn to produce $y^t$ whenever the trigger is present.
This goal can be formulated as
$
\arg \min_\delta \sum_{(x, y) \in \mathbf{D}} \mathcal{L} (\mathbf{M} (\mathcal{T} (x, \delta)), y^t),
$
where $\mathcal{L}(\cdot)$ is the loss function for the model prediction on $x^t$ and the target output $y^t$, and $y^t$ is constructed from $y$.

Directly optimizing this objective is impractical because $\mathbf{M}$ is unknown to the attacker.
Following prior studies~\cite{zengNarcissusPracticalCleanLabel2023,lanFlowMurStealthyPractical2024}, we instead optimize the trigger on a surrogate dataset $\mathbf{D}_s$ and a surrogate model $\mathbf{S}$ obtained in Stage 1:
\begin{equation}
\arg \min_\delta \sum_{(x, y) \in \mathbf{D}_s} \mathcal{L} (\mathbf{S} (\mathcal{T} (x, \delta)), y^t).
\label{eq:sur_obj}
\end{equation}

\noindent \textbf{Trigger Parameters.}
We represent $\delta$ using a parameter vector $\boldsymbol{\lambda}$ that includes the following attributes:
\textit{(1) Shape}: the geometric structure and visual appearance of triggers (e.g., shadow-like geometric forms or more complex designs), designed for seamless integration into the environment.
\textit{(2) Size}: the spatial extent of the trigger.
Larger triggers are typically more reliable but less stealthy.
\textit{(3) Position}: the trigger location in the image, which affects attack outcomes due to the structured, per-pixel semantic predictions.
\textit{(4) Quantity}: the number of triggers inserted.
Adding more triggers can make the attack more reliable, but also less stealthy.
\textit{(5) Intensity}: the trigger strength or transparency,
where lower intensity improves stealth at the cost of reduced attack reliability.
For each attribute $\lambda_p \in \boldsymbol{\lambda}$, we define a search space consisting of a set of predefined options.
For example, the shape attribute includes \emph{circle}, \emph{square}, \emph{triangle}, and a \emph{Batman-logo}.
We provide the complete list of these options in~\Cref{sec:badseg_additional}.

\noindent \textbf{Discrete Parameters.}
Since $\delta=\mathcal{F}(\boldsymbol{\lambda})$,
optimizing $\delta$ reduces to finding the best $\boldsymbol{\lambda}$ that minimizes~\Cref{eq:sur_obj}.
A practical way is to adopt a differentiable approach, which enables gradient-based optimization.
However, finding the best set of $\boldsymbol{\lambda}$ is challenging because each attribute is selected from a discrete candidate set, making the objective non-differentiable.
In particular, hard argmax selections or discrete sampling introduce discontinuities, blocking gradient backpropagation.
To address this issue, we employ the Gumbel-Softmax reparameterization~\cite{jangCategoricalReparameterizationGumbelSoftmax2017,maddisonConcreteDistributionContinuous2017}.

For each parameter $\lambda_p$, we use Gumbel-Softmax reparameterization to obtain a soft one-hot selection $\boldsymbol{\eta}^{(p)}$ through the following equation:
\begin{equation}
\eta^{(p)}_i=
\frac{\exp\left((\mathcal{G}^{(p)}_i+\log \xi^{(p)}_i)/\tau\right)}
{\sum_j \exp\left((\mathcal{G}^{(p)}_j+\log \xi^{(p)}_j)/\tau\right)},
\label{eq:gumbel}
\end{equation}
where $\boldsymbol{\xi}^{(p)}$ is a set of categorical probabilities for $\lambda_p$ and $\mathcal{G}^{(p)}$ are i.i.d.\ Gumbel noises.
$\tau$ is a temperature parameter.
We progressively lower $\tau$ to obtain near one-hot samples of $\boldsymbol{\eta}^{(p)}$.

With $\boldsymbol{\eta}^{(p)}$, we obtain a relaxed parameter $\tilde{\lambda}_p=\phi_p(\boldsymbol{\eta}^{(p)})$.
We aggregate all relaxed parameters as $\tilde{\boldsymbol{\lambda}}$.
This yields a differentiable trigger $\delta=\mathcal{F}(\tilde{\boldsymbol{\lambda}})$ and enables end-to-end optimization on
$\{\boldsymbol{\xi}^{(p)}\}_{p=1}^k$ for the surrogate objective:
\begin{equation}
\min_{\{\boldsymbol{\xi}^{(p)}\}_{p=1}^k}\;
\sum_{(x, y)\in\mathbf{D}_s}
\mathcal{L}\Big(
\mathbf{S}\big(\mathcal{T}(x,\mathcal{F}(\tilde{\boldsymbol{\lambda}}))\big),\, y^t
\Big).
\label{eq:sur_gs}
\end{equation}
At each iteration, we sample a single set of Gumbel noises to construct $\boldsymbol{\eta}^{(p)}$.
After optimization in~\Cref{eq:sur_gs}, we discretize each parameter by taking the most probable choice under $\boldsymbol{\xi}^{(p)}$ and construct the final trigger $\delta$.
More details are presented in~\Cref{sec:badseg_additional}.

\subsection{Stage 3: Label Manipulation}

This stage selects effective victim--target label pairs for constructing targeted segmentation backdoors.
We determine these pairs based on inter-class semantic distance, since it directly influences attack performance.
Intuitively, mapping between semantically distant classes (e.g., \emph{road} to \emph{sky}) typically requires stronger poisoning signals.
In contrast, mapping between similar classes (e.g., \emph{road} to \emph{sidewalk}) requires less aggressive poisoning.

\noindent \textbf{Class Center Calculation.}
Inspired by prior work~\cite{huangCARClassAwareRegularizations2022,yuanObjectContextualRepresentationsSemantic2020,zhangACFNetAttentionalClass2019}, we calculate a \emph{class center} for each category in feature space by measuring their semantic distance.
Given an input image $x$, the surrogate model $\mathbf{S}$ produces a feature map $\mathbf{F}\in\mathbb{R}^{C\times H\times W}$ and a one-hot label map $\mathbf{Y}\in\mathbb{R}^{K\times H\times W}$ from ground truth labels.
We flatten them to $\mathbf{F}_{\text{flat}}\in\mathbb{R}^{HW\times C}$ and $\mathbf{Y}_{\text{flat}}\in\mathbb{R}^{HW\times K}$.
We then define a class-center matrix $\mu\in\mathbb{R}^{K\times C}$ as
$
\mu = (\mathbf{Y}_{\text{flat}}^T \mathbf{F}_{\text{flat}}) / \nu,
$
where $\nu$ denotes the number of non-zero pixels in $\mathbf{Y}_{\text{flat}}$.

\noindent \textbf{Label Selection.}
Given $\mu$, we quantify similarity between classes $i$ and $j$ using the Euclidean distance
$d(\mu_i,\mu_j)=\|\mu_i-\mu_j\|_2$.
Smaller distances indicate higher feature-level similarity, making such pairs suitable for stealthier manipulations, whereas larger distances typically require more poisoning effort.
We use this metric to select victim--target pairs.

\noindent \textbf{Global Averaging.}
To reduce variance across images, we aggregate class centers over the surrogate dataset.
Specifically, we compute a center matrix for each mini-batch and average them to obtain a global center matrix $\bar{\mu}$.
We then compute pairwise distances using $\bar{\mu}$, i.e., $d(\bar{\mu}_i,\bar{\mu}_j)=\|\bar{\mu}_i-\bar{\mu}_j\|_2$, and select pairs accordingly.

By targeting victim--target pairs with minimal semantic distance, we exploit their feature similarity to ensure more efficient attacks.

\subsection{Backdoor Implementation}
Lastly, the adversary inserts a backdoor into the target model $\mathbf{M}$ by training it on a poisoned dataset that contains both the clean and triggered samples.
With the  clean subset $\mathbf{D}_c$ and the triggered subset $\mathbf{D}_t$, $\mathbf{D} = \mathbf{D}_c \cup \mathbf{D}_t$.
Now $\mathbf{D}$ is a poisoned dataset with both clean and triggered samples.
The model parameters $\theta$ are learned by minimizing the combined loss:
\begin{equation*}
\mathcal{L}_\text{total} =
\sum_{(x,y)\in \mathbf{D}_c}\mathcal{L}\!\left(\mathbf{M}(x;\theta),y\right)
\;+\;
\sum_{(x^t,y^t)\in \mathbf{D}_t}\mathcal{L}\!\left(\mathbf{M}(x^t;\theta),y^t\right)
,
\end{equation*}
where $\mathcal{L}(\cdot)$ denotes the task loss.

Training on $\mathbf{D}$ yields a backdoored model that maintains high accuracy on clean inputs,
but outputs the attacker-chosen target label whenever the trigger is present in the input.

\noindent \textbf{Applying BADSEG for Backdoor Attacks.}
BADSEG supports all proposed attack vectors.
Coarse-grained attacks are specified by the chosen victim and target classes, whereas fine-grained attacks additionally impose instance-level constraints or activation conditions.
Given an attack vector, BADSEG follows a unified pipeline: it first optimizes trigger parameters on a surrogate model, poisons a small subset of training samples by implanting the optimized trigger, and then modifies the target pixel labels accordingly.
This design enables BADSEG to conduct diverse segmentation backdoor attacks with consistent procedures.

\section{Evaluation}
\label{sec:evaluation}

\subsection{Experimental Setup}
\label{sec:exp_setup}

\noindent \textbf{Datasets and Models.}
We evaluate BADSEG on BDD100K~\cite{yuBDD100KDiverseDriving2020} and Cityscapes~\cite{cordtsCityscapesDatasetSemantic2016}, two widely adopted autonomous-driving benchmarks.
We consider three representative segmentation models: PSPNet~\cite{zhaoPyramidSceneParsing2017}, DeepLabV3~\cite{chenEncoderDecoderAtrousSeparable2018}, and ConvNeXt-T with the UPerNet head~\cite{liuConvNet2020s2022}.

\noindent \textbf{Evaluation Metrics.}
We consider the following evaluation metrics:
\begin{itemize}[noitemsep, topsep=0pt]
\item \emph{Attack effectiveness:} Attack Success Rate (\textit{ASR}) for poisoned data (\(\uparrow\)).
\item \emph{Model utility:} Clean Benign Accuracy (\textit{CBA}) for clean data (\(\uparrow\)) and Poisoned Benign Accuracy (\textit{PBA}) for poisoned data (\(\uparrow\)).
\item \emph{Attack stealthiness:} PSNR ($\uparrow$), SSIM ($\uparrow$), and LPIPS ($\downarrow$) between clean and poisoned data to quantify trigger imperceptibility~\cite{jiangColorBackdoorRobust2023}.
\end{itemize}
More details are presented in~\Cref{sec:exp_setup_additional}.

\noindent \textbf{Attacks.} We evaluate following attacks: Object-to-Object Attacks (\textbf{O2O}), Object-to-Background Attacks (\textbf{O2B}), Background-to-Object Attacks (\textbf{B2O}), Background-to-Background Attacks (\textbf{B2B}), Instance-Level Attacks (\textbf{INS}), and Conditional Attacks (\textbf{CON}).

\begin{table}[t]
\scriptsize
  \centering
  \caption{Optimized trigger parameters for coarse-grained attacks.}
    \begin{tabular}{c|ccccc}
    \toprule
    \textbf{Attack} & \textbf{Shape} & \textbf{Size} & \textbf{Position} & \textbf{Quantity} & \textbf{Intensity} \\
    \midrule
    O2O & Triangle  & 1/8   & Obj center    & 1 & 0.6 \\
    O2B & Circle    & 1/8   & Obj center    & 1 & 0.4 \\
    B2O & Circle    & 1/8   & -             & 1 & 0.4 \\
    B2B & Triangle  & 0.025 & -             & 10    & 0.4 \\
    \bottomrule
    \end{tabular}%
\label{tab:exp_trigger_design}%
\end{table}%

\begin{figure}[t]
  \scriptsize
  \centering
  \includegraphics[width=\linewidth]{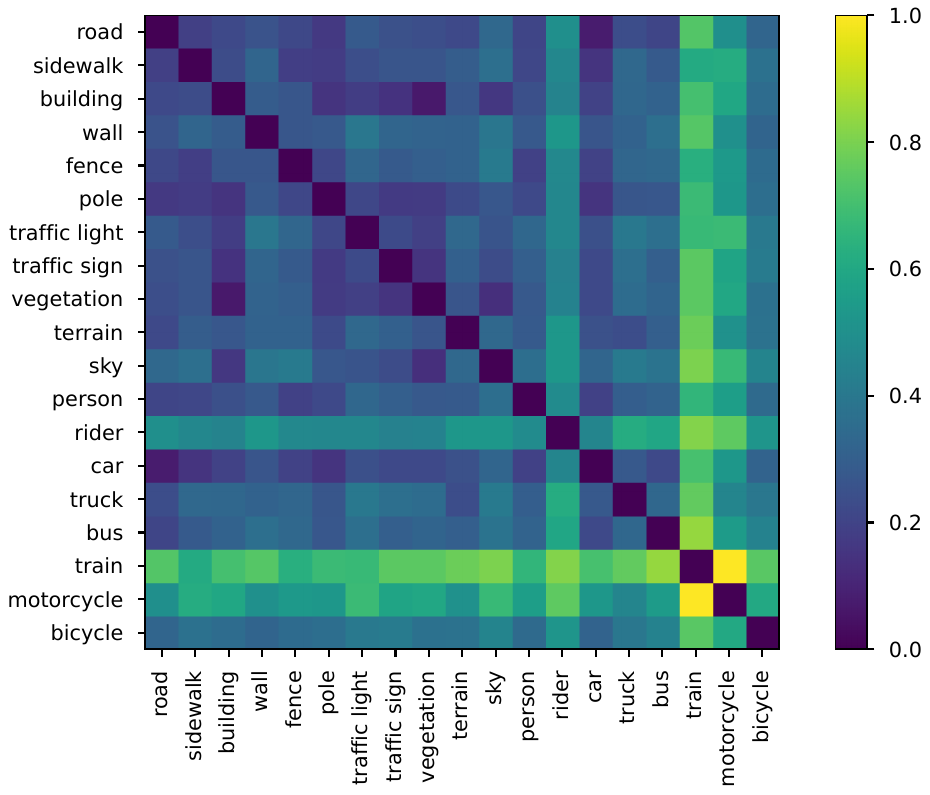}
  \caption{Normalized distance matrix across class pairs.
  Smaller values indicate stronger semantic correlations.}
  \label{fig:class_center_matrix}
\end{figure}

\noindent \textbf{Optimized Trigger Parameters.}
\Cref{tab:exp_trigger_design} reports the optimized trigger parameter results for each coarse-grained attack.
We observe that the results differ across attack vectors.
For instance, O2O prefers a \emph{triangle} trigger with higher intensity (0.6), whereas O2B and B2O converge to a \emph{circle} with lower intensity (0.4).
In contrast, B2B favors a much smaller trigger (0.025) with a higher quantity (10).
These differences suggest that trigger design is strongly coupled with the underlying attack vector.
We further analyze the impact of different parameter choices in the ablation studies.
In the following experiments, we adopt the results reported in \Cref{tab:exp_trigger_design} for each attack.

\begin{table}[t]
\scriptsize
  \centering
  \caption{Top 20 closest class pairs by normalized distance. }
    \begin{tabular}{cc|cc}
    \toprule
    \textbf{Rank} & \textbf{Class Pair} & \textbf{Distance} & \textbf{Suitable Attacks} \\
    \midrule
    1 &(building, vegetation) & 0.0666 & B2B \\
    2 &(car, road) & 0.0744 & O2B, B2O \\
    3 &(sky, vegetation) & 0.1343 & B2B \\
    4 &(building, traffic sign) & 0.1476 & B2B \\
    5 &(car, pole) & 0.1506 & O2B, B2O \\
    6 &(pole, building) & 0.1509 & B2B \\
    7 &(traffic sign, vegetation) & 0.1510 & B2B \\
    8 &(sidewalk, car) & 0.1517 & O2B, B2O \\
    9 &(building, sky) & 0.1622 & B2B \\
    10 &(pole, road) & 0.1673 & O2B, B2O \\
    11 &(pole, traffic sign) & 0.1705 & B2B \\
    12 &(pole, vegetation) & 0.1743 & B2B \\
    13 &(pole, sidewalk) & 0.1786 & B2B \\
    14 &(traffic light, building) & 0.1802 & B2B \\
    15 &(sidewalk, fence) & 0.1837 & B2B \\
    16 &(traffic light, vegetation) & 0.1880 & B2B \\
    17 &(sidewalk, road) & 0.1889 & B2B \\
    18 &(person, car) & 0.1924 & O2O \\
    19 &(fence, person) & 0.1935 & O2B, B2O \\
    20 &(car, building) & 0.1959 & O2B, B2O \\
    \bottomrule
    \end{tabular}%
  \label{tab:top20_class_pairs}%
\end{table}%

\noindent \textbf{Victim--Target Pair Selection.}
We select victim--target pairs by measuring inter-class distances in the feature space.
\Cref{fig:class_center_matrix} shows the normalized distance matrix, where smaller values indicate higher semantic similarity.
The matrix shows clear clustering patterns:
\emph{stuff} categories are generally closer to one another than \emph{object} categories, indicating higher feature similarity within stuff classes.
As a result, the closest \emph{stuff} pairs (e.g., \emph{building}--\emph{vegetation}) are ideal for the proposed B2B attacks.
By contrast, many \emph{object} categories (e.g., \emph{rider} and \emph{train}) are more isolated and lie farther from most other classes.
However, some \emph{object} pairs remain highly similar (e.g., \emph{person}--\emph{car}), making them suitable for O2O attacks.
These results provide a principled criterion for choosing effective victim--target pairs across attack vectors.

\Cref{tab:top20_class_pairs} lists the top 20 closest pairs, which are dominated by B2B pairs.
Guided by the ranking, we set our attacks using: (car $\rightarrow$ person) for O2O, (car $\rightarrow$ road) for O2B, (road $\rightarrow$ car) for B2O, and (sidewalk $\rightarrow$ road) for B2B.

\begin{table}[t]
\scriptsize
  \centering
  \caption{Results of coarse-grained backdoor attacks.}
    \begin{tabular}{c|c|ccc}
    \toprule
    \textbf{Attack} & \textbf{Model} & \textbf{ASR $\uparrow$} & \textbf{PBA $\uparrow$} & \textbf{CBA $\uparrow$} \\
    \midrule
          & ConvNeXt-T  & 0.9352 & 0.6090 & 0.6265  \\
    O2O   & PSPNet      & 0.9088 & 0.4814 & 0.5053  \\
          & DeepLabV3   & 0.9371 & 0.5411 & 0.5743  \\
    \midrule
          & ConvNeXt-T  & 0.9428 & 0.6283 & 0.6310  \\
    O2B   & PSPNet      & 0.9330 & 0.5703 & 0.5784  \\
          & DeepLabV3   & 0.9098 & 0.5708 & 0.5790  \\
    \midrule
          & ConvNeXt-T  & 0.9835 & 0.6159 & 0.6260  \\
    B2O   & PSPNet      & 0.9781 & 0.4779 & 0.4976  \\
          & DeepLabV3   & 0.9592 & 0.6015 & 0.5873 \\
    \midrule
          & ConvNeXt-T  & 0.9315 & 0.6281 & 0.6282  \\
    B2B   & PSPNet      & 0.9316 & 0.5033 & 0.4909  \\
          & DeepLabV3   & 0.9052 & 0.5796 & 0.5743  \\
    \bottomrule
    \end{tabular}%
  \label{tab:overall_performance}%
\end{table}%

\subsection{Effectiveness of Coarse-Grained Attacks}
\label{sec:exp_main_results}

\noindent \textbf{Attack results.}
\Cref{tab:overall_performance} reports results of ConvNeXt-T, PSPNet, and DeepLabV3 under four coarse-grained attack vectors on BDD100K.
On BDD100K, all three models are highly vulnerable: ASRs remain consistently high across vectors (0.91--0.98).
ConvNeXt-T and DeepLabV3 generally achieve stronger attacks than PSPNet, while maintaining higher model utility (PBA/CBA).
Across vectors, B2O is the most effective, yielding the highest ASRs for all models.
In terms of model utility, PBA and CBA vary across backbones and attack vectors.
Yet, all results remain comparable to the clean sample segmentation results reported in~\Cref{tab:model_utility}, indicating that these attacks largely preserve the victim model utility.

\begin{table}[t]
\scriptsize
  \centering
  \caption{Segmentation results on clean data.}
    \begin{tabular}{c|cc}
    \toprule
    \textbf{Model} & \textbf{BDD100K} & \textbf{Cityscapes} \\
    \midrule
    ConvNeXt-T & 0.6321 & 0.7832 \\
    PSPNet     & 0.5888 & 0.7285 \\
    DeepLabV3  & 0.6062 & 0.7661 \\
    \bottomrule
    \end{tabular}%
  \label{tab:model_utility}%
\end{table}%

\begin{table}[t]
\scriptsize
  \centering
  \caption{Rank correlation of class pairs across segmentation architectures.}
        \resizebox{\linewidth}{!}{
        \begin{tabular}{c|ccccc}
    \toprule
    \multirow{2}[4]{*}{\textbf{Models}} & \textbf{Spearman's } & \textbf{Kendall's } & \multicolumn{3}{c}{\textbf{Top-$K$ overlap (\#)}} \\
\cmidrule{4-6}          & \textbf{Rank Corr} & \textbf{Tau} & \textbf{20} & \textbf{50} & \textbf{100} \\
    \midrule
    ConvNeXt-T, PSPNet & 0.7921 & 0.6011 & 13    & 32    & 83 \\
    ConvNeXt-T, DeepLabV3 & 0.7831 & 0.5915 & 13    & 32    & 83 \\
    PSPNet, DeepLabV3 & 0.9982 & 0.9696 & 19    & 50    & 100 \\
    \bottomrule
    \end{tabular}%
    }
  \label{tab:ablation_class_pair_simi}%
\end{table}%

\subsection*{Ablation Study on Stage 1 Backdoor Preparation}

We conduct attacks considering different surrogate models.

\noindent \textbf{Impact of Surrogate Models.}
To examine the impact of the selected surrogate model on trigger parameter selection, we evaluate segmentation models of PSPNet, DeepLabV3, and ConvNeXt-T.
Our evaluation shows that the selected parameters are largely consistent with~\Cref{tab:exp_trigger_design}: the trigger typically adopts a triangle or circle shape with size $1/8$, is placed at the object center, uses a single instance, and selects an intensity of $0.6$ or $0.4$.
The consistency suggests that our trigger parameter optimization is robust across different surrogate model architectures.

To examine the impact of the selected surrogate models on victim--target pair selections, we recompute class-pair distance rankings for each surrogate architecture and report their correlation results in~\Cref{tab:ablation_class_pair_simi}.
The table shows high rank correlations across surrogate architectures with considerable overlapped Top-$K$ victim-target pairs.
In particular, PSPNet and DeepLabV3 yield near-identical rankings, while ConvNeXt-T also aligns closely.
Similarly, this result indicates that the calculated victim-target pair ranking is robust across surrogate models.

The ablation study demonstrates that attackers can adopt surrogate segmentation models that are different from the victim models to launch effective backdoor attacks.

\begin{table}[t]
\scriptsize
  \centering
  \caption{Impact of trigger shape and size on attack performance.}
    \resizebox{\linewidth}{!}{
    \begin{tabular}{c|lccc||lccc}
    \toprule
    \multirow{2}[4]{*}{\textbf{Attack}} & \multicolumn{4}{c||}{\textbf{Various Shapes (fixed size)}} & \multicolumn{4}{c}{\textbf{Various Sizes (fixed shape)}} \\
\cmidrule{2-9}          & \textbf{Shape} & \textbf{ASR $\uparrow$} & \textbf{PBA $\uparrow$} & \textbf{CBA $\uparrow$} & \textbf{Size} & \textbf{ASR $\uparrow$} & \textbf{PBA $\uparrow$} & \textbf{CBA $\uparrow$} \\
    \midrule
          & Circle   & 0.9352           & 0.6090 & 0.6265 & 1/12  & 0.9373          & 0.6081 & 0.6326 \\
          & Square   & 0.9284           & 0.6166 & 0.6329 & 1/10  & 0.9249          & 0.6164 & 0.6379 \\
    O2O   & Triangle & \textbf{0.9368}  & 0.5970 & 0.6234 & 1/8   & \textbf{0.9452} & 0.6090 & 0.6265 \\
          & Logo     & 0.9363           & 0.6066 & 0.6291 & 1/6   & 0.9154          & 0.6127 & 0.6289 \\
          &         &                 &       &       & 1/4   & 0.9397          & 0.6119 & 0.6349 \\
    \midrule
          & Circle   & 0.9428           & 0.6283 & 0.6310 & 1/12  & 0.9230          & 0.6255 & 0.6288 \\
          & Square   & 0.9195           & 0.6315 & 0.6339 & 1/10  & 0.9254          & 0.6257 & 0.6331 \\
    O2B   & Triangle & 0.9190           & 0.6294 & 0.6379 & 1/8   & 0.9428          & 0.6283 & 0.631 \\
          & Logo     & \textbf{0.9464}  & 0.6326 & 0.6362 & 1/6   & \textbf{0.9516} & 0.6257 & 0.6276 \\
          &         &                 &       &       & 1/4   & 0.9357          & 0.6276 & 0.6334 \\
    \midrule
          & Circle   & \textbf{0.9835}  & 0.6159 & 0.6260 & 1/12  & 0.9789          & 0.6151 & 0.6285 \\
          & Square   & 0.9742           & 0.6198 & 0.6287 & 1/10  & 0.9812          & 0.6139 & 0.6271 \\
    B2O   & Triangle & 0.9789           & 0.6141 & 0.6315 & 1/8   & \textbf{0.9835} & 0.6159 & 0.626 \\
          & Logo     & 0.9821           & 0.6173 & 0.6298 & 1/6   & 0.9793          & 0.6115 & 0.6247 \\
          &         &                 &       &       & 1/4   & 0.9721          & 0.6098 & 0.6234 \\
    \midrule
          & Circle   & 0.9315           & 0.6281 & 0.6282 & 0.005 & 0.9141          & 0.6278 & 0.6285 \\
          & Square   & 0.9235           & 0.6265 & 0.6256 & 0.010 & 0.9221          & 0.6332 & 0.6369 \\
    B2B   & Triangle & \textbf{0.9360}  & 0.6342 & 0.6330 & 0.015 & \textbf{0.9315} & 0.6281 & 0.6282 \\
          & Logo     & 0.9342           & 0.6295 & 0.6299 & 0.020 & 0.9265          & 0.6296 & 0.6353 \\
          &         &                 &       &       & 0.025 & 0.9270          & 0.6311 & 0.631 \\
    \bottomrule
    \end{tabular}%
    }
  \label{tab:exp_trigger_shape_size}%
\end{table}

\begin{table}[t]
\scriptsize
\centering
  \caption{Impact of trigger positions on attack performance.}
    \begin{tabular}{c|c|ccc}
    \toprule
    \textbf{Attack} & \textbf{Position} & \textbf{ASR $\uparrow$} & \textbf{PBA $\uparrow$} & \textbf{CBA $\uparrow$} \\
    \midrule
        & Object center         & \textbf{0.9352} & 0.6090 & 0.6265 \\
    O2O & Random on object      & 0.9282 & 0.6126 & 0.6346 \\
        & Random outside object & 0.9020 & 0.6092 & 0.6356 \\
    \midrule
        & Object center         & \textbf{0.9428} & 0.6283 & 0.6310 \\
    O2B & Random on object      & 0.8885 & 0.6261 & 0.6120 \\
        & Random outside object & 0.7014 & 0.6251 & 0.6345 \\
    \bottomrule
    \end{tabular}%
  \label{tab:exp_trigger_position}%
\end{table}%

\begin{table}[t]
\scriptsize
  \centering
  \caption{Impact of trigger quantity and intensity on attack performance.}
      \resizebox{\linewidth}{!}{
    \begin{tabular}{c|cccc||cccc}
    \toprule
    \multirow{2}[4]{*}{\textbf{Attack}} & \multicolumn{4}{c||}{\textbf{Trigger Quantity Results}} & \multicolumn{4}{c}{\textbf{Trigger Intensity Results}} \\
\cmidrule{2-9}          & \textbf{Quan} & \textbf{ASR $\uparrow$} & \textbf{PBA $\uparrow$} & \textbf{CBA $\uparrow$} & \textbf{Inten} & \textbf{ASR $\uparrow$} & \textbf{PBA $\uparrow$} & \textbf{CBA $\uparrow$} \\
    \midrule
          & 1     & \textbf{0.9352} & 0.6090 & 0.6265 & 0.2   & 0.8940          & 0.6047 & 0.6290 \\
    O2O   & 3     & 0.9162          & 0.6074 & 0.6315 & 0.4   & 0.9352          & 0.6090 & 0.6265 \\
          & 5     & 0.9087          & 0.6058 & 0.6358 & 0.6   & \textbf{0.9392} & 0.6110 & 0.6296 \\
          &      &                &       &       & 0.8   & 0.9373          & 0.6172 & 0.6425 \\
    \midrule
          & 1     & \textbf{0.9428} & 0.6283 & 0.6310 & 0.2   & 0.8937          & 0.6296 & 0.6287 \\
    O2B   & 3     & 0.9281          & 0.6235 & 0.6304 & 0.4   & 0.9428          & 0.6283 & 0.6310 \\
          & 5     & 0.9145          & 0.6194 & 0.6287 & 0.6   & 0.9274          & 0.6151 & 0.6262 \\
          &      &                &       &       & 0.8   & \textbf{0.9464} & 0.6229 & 0.6274 \\
    \midrule
          & 1     & \textbf{0.9835} & 0.6159 & 0.6260 & 0.2   & 0.9524          & 0.6089 & 0.6198 \\
    B2O   & 3     & 0.9718          & 0.6098 & 0.6285 & 0.4   & 0.9835          & 0.6159 & 0.6260 \\
          & 5     & 0.9612          & 0.6071 & 0.6308 & 0.6   & 0.9813          & 0.6125 & 0.6245 \\
          &      &                &       &       & 0.8   & \textbf{0.9896} & 0.6148 & 0.6268 \\
    \midrule
          & 1     & 0.4864          & 0.6275 & 0.6232 & 0.2   & 0.9083          & 0.6375 & 0.6351 \\
    B2B   & 5     & 0.9315          & 0.6281 & 0.6282 & 0.4   & 0.9315          & 0.6281 & 0.6282 \\
          & 10    & \textbf{0.9488} & 0.6314 & 0.6309 & 0.6   & 0.9246          & 0.6398 & 0.6394 \\
          &      &                &       &       & 0.8   & \textbf{0.9364} & 0.6347 & 0.6289 \\
    \bottomrule
    \end{tabular}%
    }
  \label{tab:exp_trigger_quantity_intensity}%
\end{table}

\subsection*{Ablation Study on Stage 2 Trigger Optimization}
In this part, we evaluate the proposed attacks with different trigger parameters.

\noindent \textbf{Impact of Trigger Shape.}
\Cref{tab:exp_trigger_shape_size} presents an ablation study with the trigger shape while keeping other parameters fixed.
Overall, the proposed attacks are insensitive to the specific shape choice: all four shapes (circle, square, triangle, and a logo-style pattern) achieve comparably high ASRs with similar PBA/CBA, indicating that the attacks are robust across different trigger shapes.

Interestingly, the effectiveness of BADSEG does not increase with trigger complexity.
Intuitively, complex patterns might result in stronger attacks.
Yet, our results show that simple geometric shapes (e.g., triangle and circle) achieve ASRs comparable to, and sometimes higher than, the more complex Logo design.
For instance, the triangle trigger achieves the highest ASR in both O2O and B2B attacks.

\noindent \textbf{Impact of Trigger Size.}
Prior results adopt a fixed trigger size derived from the trigger optimization.
Here, we evaluate a broader range of relative sizes to assess their impact on attack performance.
We represent the trigger size relative to the scene scale: for B2B, the size is defined as a fraction of the image width; for object-based vectors, it is defined as a fraction of the target object width.
\Cref{tab:exp_trigger_shape_size} reveals a non-monotonic relationship between trigger size and attack results: the
performance generally peaks at a relative size of $1/8$ (or $0.015$ for B2B), after which increasing the size leads to slight degradation.
These finding suggests a trade-off: the trigger must be large enough to provide a reliable trigger signal, yet small enough to avoid perturbing the global semantic context.

\noindent \textbf{Impact of Trigger Position.}
We evaluate the Impact of trigger position for O2O and O2B under three placement strategies: (i) at the target object center, (ii) at a random location within the target object, and (iii) at a random location outside the target object.
\Cref{tab:exp_trigger_position} presents the results.
For O2O, the attack remains robust to relocation: moving the trigger from the object center to an on-object random position only slightly reduces ASR (0.9352 $\rightarrow$ 0.9282), and even placing it outside the object leads to only a marginal drop ( 0.9020).
In contrast, O2B is highly position-sensitive: while center placement achieves the highest ASR (0.9428), shifting the trigger to a random on-object location reduces ASR to 0.8885, and placing it outside the object causes a decrease to 0.7014.
These results suggest that O2B works best when the trigger is on the target object, while O2O is relatively insensitive to trigger positions.

\noindent \textbf{Impact of Trigger Quantity.}
We study the effect of trigger quantity with different numbers of triggers per image.
For object-targeted vectors, we test ${1,3,5}$ triggers; for B2B, we set a wider range ${1,5,10}$ as stuff regions are usually larger.
\Cref{tab:exp_trigger_quantity_intensity} presents the attack results.
The results show that adding triggers improves ASR for B2B.
We attribute this gain to improved spatial coverage: distributing more triggers across the image better spans the large stuff regions targeted in B2B attacks.
In contrast, for object-targeted vectors, adding triggers yields marginal gains and can even slightly reduce ASR, indicating that a single trigger is usually sufficient.

\noindent \textbf{Impact of Trigger Intensity.}
We vary the trigger intensity from $0.2$ to $0.8$ and report the results in \Cref{tab:exp_trigger_quantity_intensity}.
Larger intensities introduce stronger (and more visible) perturbations, whereas smaller intensities yield stealthier triggers.
Overall, increasing intensity tends to improve ASR across attack vectors.
However, it comes at the cost of reduced visual stealthiness, reflecting a clear effectiveness--stealth trade-off.
Notably, moderate intensities (e.g., $0.4$--$0.6$) already achieve high ASRs comparable to the strongest setting, while allowing the trigger to remain well blended with surrounding pixels.

\noindent \textbf{Impact on Model Utility.}
\Cref{tab:exp_trigger_shape_size,tab:exp_trigger_position,tab:exp_trigger_quantity_intensity} demonstrate that the segmentation models maintain robust performance across proposed attacks.
Both PBA and CBA exhibit minor variations within a consistent range (approximately 0.59--0.64 mIoU).
They remain comparable to the clean-sample segmentation results reported in~\Cref{tab:model_utility}.
This suggests that variations in trigger parameters primarily determine the efficacy of backdoor activation, while having a negligible impact on the utility of the victim model.

\begin{table}[t]
\centering
\scriptsize
\caption{Attack performance under various victim--target class configurations for O2B attacks.}
\label{tab:ablation_poisoning_o2b}

\begin{subtable}[t]{0.23\textwidth}
\centering
\caption{Varying target classes.}
\label{tab:poisoning_o2b_victim_unchanged_sub}
\begin{tabular}{l|l|c}
\toprule
\textbf{Victim} & \textbf{Target} & \textbf{ASR $\uparrow$}  \\
\midrule
\multirow{11}{*}{car} & road          & \textbf{0.9428} \\
       & sidewalk      & 0.9280  \\
       & building      & 0.9288  \\
       & wall          & 0.9058  \\
       & fence         & 0.9083  \\
       & pole          & 0.9187  \\
       & traffic light & 0.9246  \\
       & traffic sign  & 0.9275  \\
       & vegetation    & 0.9427  \\
       & terrain       & 0.9154  \\
       & sky          & 0.9250  \\
\bottomrule
\end{tabular}%
\end{subtable}
\hfill
\begin{subtable}[t]{0.23\textwidth}
\centering
\caption{Varying victim classes.}
\label{tab:poisoning_o2b_target_unchanged_sub}
\begin{tabular}{l|l|c}
\toprule
\textbf{Victim} & \textbf{Target} & \textbf{ASR $\uparrow$}  \\
\midrule
person     & \multirow{8}{*}{road} & 0.7498  \\
rider      &         & 0.4054  \\
car        &         & \textbf{0.9428}  \\
truck      &         & 0.7264  \\
bus        &         & 0.7850  \\
train      &         & 0.0143  \\
motorcycle &         & 0.4051 \\
bicycle    &         & 0.7968 \\
\bottomrule
\end{tabular}%
\end{subtable}
\end{table}

\begin{figure}[t]
  \scriptsize
  \centering
  \includegraphics[width=\linewidth]{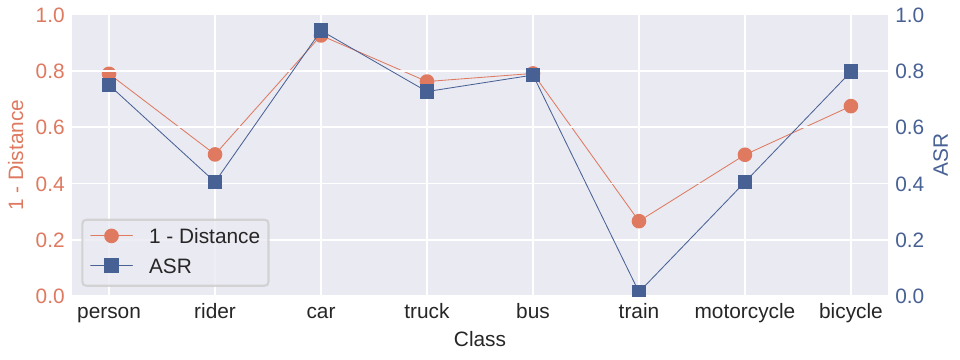}
  \caption{Euclidean distance of victim--target pairs and ASRs for O2B Attacks in~\Cref{tab:poisoning_o2b_target_unchanged_sub}.}
  \label{fig:class_vs_distance_asr}
\end{figure}

\subsection*{Ablation Study on Stage 3 Label Manipulation}
We now explore how the selection of victim--target pairs influences attack performance.

\noindent \textbf{Impact of Target Class.}
\Cref{tab:poisoning_o2b_victim_unchanged_sub} evaluates O2B attack performance using a fixed victim class (\emph{car}) paired with various stuff target classes.
We observe consistent ASRs exceeding 0.90 across all targets, indicating the majority of \emph{car} pixels are successfully misclassified as the target label.
Notably, targeting \emph{road} achieves the highest ASR, a result that aligns with the high ranked class pairs identified in \Cref{tab:top20_class_pairs}.
This suggests that higher victim-target similarity can lead to superior attack results.
Moreover, the results confirm that attacks remain effective even when the target is semantically distant from the victim (e.g., \emph{car}$\to$\emph{sky}).

\noindent \textbf{Impact of Victim Class.}
\Cref{tab:poisoning_o2b_target_unchanged_sub} analyzes O2B attacks where various victim classes are targeted to be mis-segmented as a fixed victim (\emph{road}).
The results show that attack performance is different across classes: \emph{car} is highly vulnerable (ASR = 0.94), whereas \emph{train} remains resistant (ASR = 0.01).
This is because the \emph{train} class is underrepresented within the dataset, leading to less effective attacks.
\Cref{fig:class_vs_distance_asr} further demonstrates that ASR increases with victim--target similarity, implying that semantically similar pairs contribute to superior attack performance.

\noindent \textbf{Summary.}
Our extensive evaluations reveal that:
(1) BADSEG is robust to surrogate architecture selections, maintaining consistent attack performance across diverse models;
(2) Ablation studies demonstrate that the proposed attack is across diverse parameter settings.
Furthermore, trigger parameters determined by BADSEG consistently achieve the highest performance, validating its effectiveness.
(3) Our results indicate that victim--target pair selection contributes to superior attack performance. Specifically, the attack achieves superior performance when the target and victim show strong semantic correlation.

\begin{table}[t]
\centering
\scriptsize
\caption{Performance of fine-grained attacks.}
\label{tab:exp_fine_grained}
\begin{subtable}[t]{\linewidth}
\scriptsize
  \centering
  \caption{Instance-level attacks.}
    \begin{tabular}{c|c|ccc}
    \toprule
    \textbf{Attack} & \textbf{Instance Number} & \textbf{ASR $\uparrow$} & \textbf{PBA $\uparrow$} & \textbf{CBA $\uparrow$} \\
    \midrule
            & 1     & 0.8934 & 0.6234 & 0.6412 \\
    INS-O2O & 3     & 0.9156 & 0.5967 & 0.6089 \\
            & All   & 0.9352 & 0.6090 & 0.6265 \\
    \midrule
            & 1     & 0.9076 & 0.6421 & 0.6456 \\
    INS-O2B & 3     & 0.9245 & 0.6156 & 0.6089 \\
            & All   & 0.9428 & 0.6283 & 0.6310 \\
    \bottomrule
    \end{tabular}%
  \label{tab:instance_attack}%
\end{subtable}%
\vfill
\begin{subtable}[t]{\linewidth}
\scriptsize
  \centering
  \caption{Conditional attacks.}
    \begin{tabular}{c|c|ccc}
    \toprule
    \textbf{Attack} & \textbf{Sample Rate} & \textbf{ASR $\uparrow$} & \textbf{PBA $\uparrow$} & \textbf{CBA $\uparrow$} \\
    \midrule
            & 0.01  & 0.9523 & 0.6145 & 0.6198 \\
    CON-O2O & 0.05  & 0.9763 & 0.6292 & 0.6369 \\
            & 0.25  & 0.9921 & 0.6210 & 0.6331 \\
    \midrule
            & 0.01  & 0.8967 & 0.6089 & 0.6145 \\
    CON-O2B & 0.05  & 0.9087 & 0.6167 & 0.6221 \\
            & 0.25  & 0.9204 & 0.6284 & 0.6336 \\
    \bottomrule
    \end{tabular}%
  \label{tab:conditional_attack}%
\end{subtable}%
\end{table}%

\subsection{Effectiveness of Fine-Grained Attacks}

\noindent \textbf{Performance of Instance-Level Attacks.}
\Cref{tab:instance_attack} evaluates instance-level attacks for two vectors (INS-O2O and INS-O2B) while varying the number of targeted instances.
The results show that BADSEG can selectively activate the backdoor attack on a subset of instances, targeting a single instance can achieve a high ASR (0.89 for INS-O2O and 0.90 for INS-O2B).
As the number of targets increases to ``All'', ASR improves further to 0.94, yet the marginal gain is relatively small.
This indicates that the backdoor can precisely target determined targets without requiring widespread poisoning.
Meanwhile, PBA and CBA remain stable (above 0.6), suggesting that these attacks induce localized mis-segmentation without affecting the segmentation of the surrounding pixels.

\noindent \textbf{Performance of Conditional Attacks.}
\Cref{tab:conditional_attack} evaluates conditional attacks for two vectors (CON-O2O and CON-O2B) under varying sample rates.
The sample rate determines the fraction of poisoned training samples that satisfy the condition and contain designed triggers.
In our evaluation, we use \emph{red cars} as the condition due to the sufficient number of samples in the dataset.
The results demonstrate that conditional attacks achieve consistently high ASRs across all sample rates, with CON-O2O generally outperforming CON-O2B.
Notably, even at low sample rates, the backdoor remains reliable.
PBA and CBA also stay stable (above 0.6), confirming that conditional attacks preserve model utility while achieving strong backdoor performance.

\begin{table}[t]
\scriptsize
  \centering
  \caption{Stealthiness comparison across various attacks.}
\resizebox{\linewidth}{!}{
    \begin{tabular}{c|ccc|c|ccc}
    \toprule
    \textbf{Attack} & \textbf{PSNR $\uparrow$} & \textbf{LPIPS $\downarrow$} & \textbf{SSIM $\uparrow$} & \textbf{Attack} & \textbf{PSNR $\uparrow$} & \textbf{LPIPS $\downarrow$} & \textbf{SSIM $\uparrow$} \\
    \midrule
    HBA  & 10.74 & 0.2319 & 0.7332 & B2O  & 35.03 & 0.0167 & 0.9874 \\
    OFBA & 28.45 & 0.0278 & 0.9649 & B2B  & 37.35 & 0.0044 & 0.9935 \\
    IBA  & 24.53 & 0.0426 & 0.9871 & INS  & \textbf{52.27} & \textbf{0.0008} & \textbf{0.9991} \\
    O2O  & 40.12 & 0.0035 & 0.9948 & CON  & 42.68 & 0.0154 & 0.9946 \\
    O2B  & 40.44 & 0.0030 & 0.9950 & & & & \\
    \bottomrule
    \end{tabular}%
    }
  \label{tab:stealthiness}%
\end{table}

\subsection{Attack Stealthiness}
\Cref{tab:stealthiness} reports stealthiness metrics (PSNR, LPIPS, and SSIM) for all proposed attacks.
For baselines, HBA~\cite{liHiddenBackdoorAttack2021} is the least stealthy (PSNR $\approx$ 10), whereas OFBA~\cite{maoObjectfreeBackdoorAttack2023} and IBA~\cite{lanInfluencerBackdoorAttack2023} provide only limited improvements, with PSNR values still below 30.
Notably, although IBA achieves a high SSIM, its low PSNR (24.53) indicates the presence of visible pixel-level residues.
In contrast, our proposed attacks consistently achieve superior stealthiness, with PSNR $> 35$, LPIPS $< 0.02$, and SSIM $> 0.98$ across all settings.
The fine-grained INS attack achieves the highest perceptual similarity (PSNR 52.27; LPIPS 0.0008), making poisoned images visually indistinguishable from their clean counterparts.
Even our weakest case (B2O, PSNR 35.03) still outperforms the best baseline (PSNR 28.45).
Overall, these results validate the stealthiness of the proposed attacks.

\noindent \textbf{Summary.}
Both Instance-Level and Conditional attacks demonstrate that segmentation models are highly vulnerable to context-aware exploitation.
By confining the attack to selected instances or attribute-based conditions, they avoid widespread, class-wide modification.
Our results show that this fine-grained control does not come at the cost of attack performance; instead, it maintains robust ASRs while offering enhanced stealth with minimal disruption to the scene.

\begin{table}[t]
\scriptsize
  \centering
  \caption{ASR results for Fine-tuning and Pruning.}
      \resizebox{\linewidth}{!}{
      \begin{tabular}{c|c|ccc|ccc}
    \toprule
    \multirow{2}[4]{*}{\textbf{Attack}} & \multicolumn{1}{c|}{\multirow{2}[4]{*}{\textbf{Original}}} & \multicolumn{3}{c|}{\textbf{Finetuning, Clean Data}} & \multicolumn{3}{c}{\textbf{Pruning, Pruned Channels}} \\
\cmidrule{3-8}          &       & \textbf{1\%}     & \textbf{5\%}     & \textbf{10\%}    & \textbf{1\%}     & \textbf{5\%}    & \textbf{10\%} \\
    \midrule
    HBA   & 0.1815 & 0.1792 & 0.1654 & 0.1423 & 0.1801 & 0.1778 & 0.1745 \\
    OFBA  & 0.7859 & 0.7587 & 0.4892 & 0.3124 & 0.7823 & 0.7712 & 0.7787 \\
    IBA   & 0.8314 & 0.8021 & 0.5256 & 0.3867 & 0.8287 & 0.8256 & 0.8201 \\
    O2O   & 0.9352 & 0.9292 & 0.6070 & 0.4897 & 0.9395 & 0.9349 & 0.9340 \\
    O2B   & 0.9428 & 0.9054 & 0.6197 & 0.5293 & 0.9337 & 0.9289 & 0.9341 \\
    B2O   & 0.9835 & 0.9756 & 0.7234 & 0.5812 & 0.9834 & 0.9816 & 0.9798 \\
    B2B   & 0.9315 & 0.9187 & 0.6845 & 0.5634 & 0.9346 & 0.9448 & 0.9465 \\
    INS   & 0.9076 & 0.8923 & 0.6512 & 0.5178 & 0.9045 & 0.8987 & 0.8934 \\
    CON   & 0.9204 & 0.9067 & 0.6723 & 0.5389 & 0.9178 & 0.9123 & 0.9089 \\
    \bottomrule
    \end{tabular}%
    }
  \label{tab:defense_finetuning_pruning}%
\end{table}%

\section{Defenses}
\label{sec:defenses}
\noindent \textbf{Motivation.} Prior studies on backdoor attacks in semantic segmentation lack a thorough evaluation of defenses.
While backdoor defenses have been proposed~\cite{liuNeuralTrojans2017,wangNeuralCleanseIdentifying2019,liNeuralAttentionDistillation2021,taoModelOrthogonalizationClass2022,gaoBackdoorDefenseAdaptively2023,wangMMBDPostTrainingDetection2024}, they primarily focus on image classification.
Their effectiveness for semantic segmentation remains underexplored.
To bridge this gap, we aim to build the first defense benchmark by evaluating both existing and our proposed attacks against six representative defenses.
As some of these defenses are not designed for pixel-wise prediction, we first adapt them to segmentation and then evaluate them across attack methods.
Specifically, our benchmark covers Fine-tuning~\cite{shaFineTuningAllYou2022}, Pruning~\cite{liuNeuralTrojans2017,lanInfluencerBackdoorAttack2023}, ABL~\cite{liAntiBackdoorLearningTraining2021}, STRIP~\cite{gaoSTRIPDefenceTrojan2019}, TeCo~\cite{liuDetectingBackdoorsInference2023}, and Beatrix~\cite{maBeatrixResurrectionsRobust2023}.

\noindent \textbf{Evaluation Metrics.}
To be consistent with prior work, we consider the following metrics for each method.
For Fine-tuning~\cite{shaFineTuningAllYou2022}, Pruning~\cite{liuNeuralTrojans2017,lanInfluencerBackdoorAttack2023}, and ABL~\cite{liAntiBackdoorLearningTraining2021}, we utilize \textbf{ASR}, \textbf{PBA}, and \textbf{CBA}.
An effective defense minimizes ASR while preserving high PBA and CBA values to ensure model utility on clean samples.
For STRIP~\cite{gaoSTRIPDefenceTrojan2019}, TeCo~\cite{liuDetectingBackdoorsInference2023}, and Beatrix~\cite{maBeatrixResurrectionsRobust2023}, we measure their poisoned sample detection results via \textbf{ACC} (accuracy), \textbf{Recall}, \textbf{F1}, and \textbf{AUC}.
Higher scores indicate superior detection performance.

\noindent \textbf{Attacks.}
We evaluate both existing attacks (HBA~\cite{liHiddenBackdoorAttack2021}, OFBA~\cite{maoObjectfreeBackdoorAttack2023}, IBA~\cite{lanInfluencerBackdoorAttack2023}) and the proposed BADSEG attacks.
For fine-grained attacks, we consider INS-O2B and CON-O2B.
The following sections describe each defense and report its results.
We present more details and findings in~\Cref{sec:defense_additional}.

\noindent \textbf{Fine-Tuning}~\cite{shaFineTuningAllYou2022} mitigates backdoors by retraining models on clean data.
Following prior work~\cite{lanInfluencerBackdoorAttack2023}, we fine-tune backdoored models for a fixed epoch of 10 using clean subsets of 1\%, 5\%, and 10\% and summarize the results in~\Cref{tab:defense_finetuning_pruning}.
As the clean data increases, ASR consistently decreases across all attacks, indicating that fine-tuning provides partial mitigation.
However, our attacks still remain comparatively robust: under the strongest setting (10\%), they still achieve ASRs that are approximately 10--20 percentage points higher than the strongest baseline.
The results show that fine-tuning can suppress segmentation backdoor behavior.
However, compared to existing attacks, our attacks exhibit stronger resistance.

\noindent \textbf{Pruning}~\cite{liuNeuralTrojans2017,lanInfluencerBackdoorAttack2023} removes \emph{less} important neurons from backdoored models to disrupt malicious activations while maintaining model utility.
Following prior studies~\cite{liuNeuralTrojans2017,lanInfluencerBackdoorAttack2023},
we use clean samples to measure activations in the final layer, rank channels by the number of activated neurons, and zero out the least-active 1\%, 5\%, and 10\% of channels.
We report the results in~\Cref{tab:defense_finetuning_pruning}.
Overall, pruning is ineffective: ASRs remain close to the no-defense baseline across all pruning ratios.
This aligns with the results in prior studies~\cite{liuNeuralTrojans2017,lanInfluencerBackdoorAttack2023}: removing a subset of \emph{less} important neurons from the last layer does not reliably mitigate segmentation backdoors.
This might stem from the fact that, in segmentation, trigger features are distributed across spatial locations and multiple channels, making them resilient to the channel removal.

\begin{table}[t]
\centering
\scriptsize
\caption{Defense results for ABL and STRIP.}
\label{tab:defense_abl_strip}

\begin{subtable}[t]{0.22\textwidth}
\scriptsize
  \centering
  \caption{Attack with ABL.}
        \resizebox{\linewidth}{!}{
    \begin{tabular}{c|ccc}
    \toprule
    \textbf{Attack} & \textbf{ASR} & \textbf{PBA} & \textbf{CBA} \\
    \midrule
    HBA   & 0.1589 & 0.5020 & 0.5246 \\
    OFBA  & 0.5542 & 0.5605 & 0.5817 \\
    IBA   & 0.7177 & 0.5781 & 0.5872 \\
    O2O   & 0.7804 & 0.5327 & 0.5523 \\
    O2B   & 0.7750 & 0.5810 & 0.5883 \\
    B2O   & 0.8962 & 0.5036 & 0.5209 \\
    B2B   & 0.9085 & 0.5512 & 0.5507 \\
    INS   & 0.8234 & 0.5423 & 0.5634 \\
    CON   & 0.8567 & 0.5289 & 0.5498 \\
    \bottomrule
    \end{tabular}%
    }
  \label{tab:abl}%
\end{subtable}%
\hfill
\begin{subtable}[t]{0.22\textwidth}
\scriptsize
  \centering
  \caption{Detection with STRIP.}
        \resizebox{\linewidth}{!}{
    \begin{tabular}{c|ccc}
    \toprule
    \textbf{Attack} & \textbf{ACC} & \textbf{Recall} & \textbf{AUC} \\
    \midrule
    HBA     & 0.478 & 0.0257 & 0.4492 \\
    OFBA    & 0.514 & 0.0171 & 0.4824 \\
    IBA     & 0.474 & 0.0414 & 0.4590 \\
    O2O     & 0.448 & 0.1072 & 0.4263 \\
    O2B     & 0.530 & 0.0858 & 0.5017 \\
    B2O     & 0.596 & 0.0985 & 0.5173 \\
    B2B     & 0.636 & 0.0357 & 0.4877 \\
    INS     & 0.512 & 0.0923 & 0.4758 \\
    CON     & 0.547 & 0.0795 & 0.4932 \\
    \bottomrule
    \end{tabular}%
    }
  \label{tab:strip}%
\end{subtable}%
\end{table}%

\begin{figure}[t]
  \scriptsize
  \begin{subfigure}[htbp]{.155\textwidth}
    \centering
    \includegraphics[width=\linewidth]{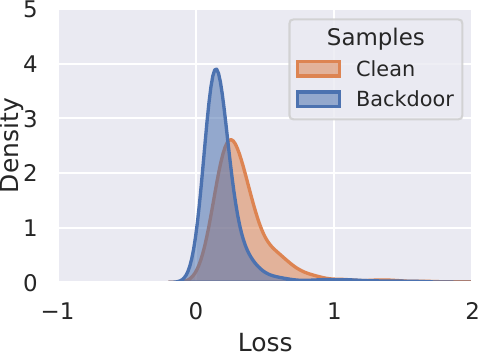}
    \caption{IBA w/ ABL}
    \label{fig:abl_kde_iba}
  \end{subfigure}
  \hfill
  \begin{subfigure}[htbp]{.155\textwidth}
    \centering
    \includegraphics[width=\linewidth]{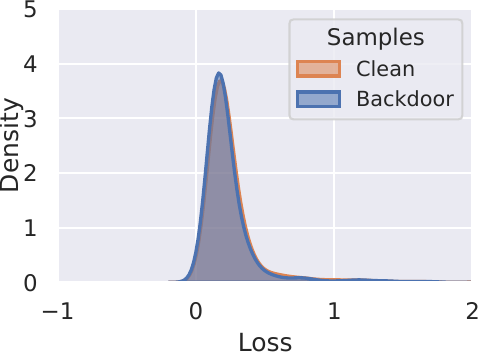}
    \caption{O2O w/ ABL}
    \label{fig:abl_kde_o2o}
  \end{subfigure}
  \hfill
  \begin{subfigure}[htbp]{.155\textwidth}
    \centering
    \includegraphics[width=\linewidth]{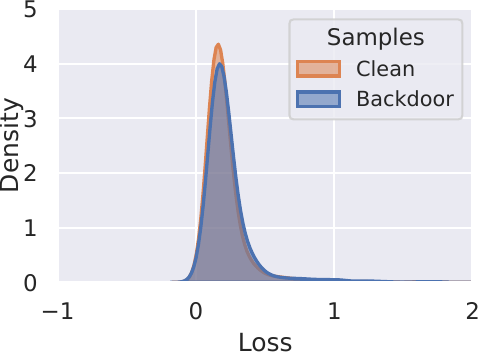}
    \caption{B2B w/ ABL}
    \label{fig:abl_kde_b2b}
  \end{subfigure}

  \begin{subfigure}[htbp]{.155\textwidth}
    \centering
    \includegraphics[width=\linewidth]{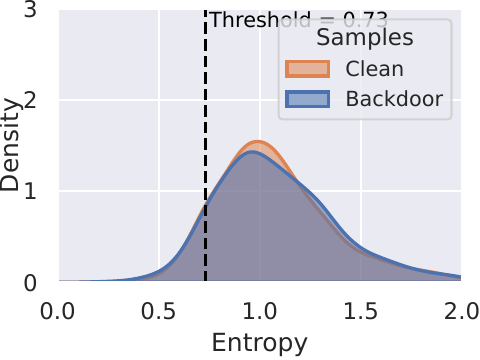}
    \caption{IBA w/ STRIP}
    \label{fig:strip_kde_iba}
  \end{subfigure}
  \hfill
  \begin{subfigure}[htbp]{.155\textwidth}
    \centering
    \includegraphics[width=\linewidth]{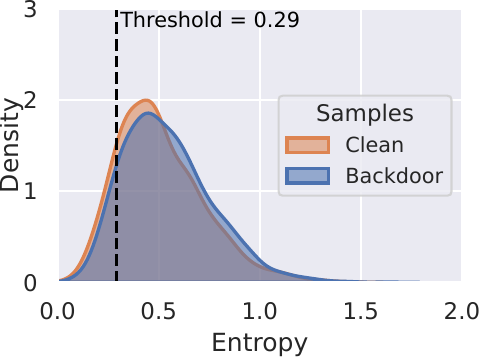}
    \caption{O2O w/ STRIP}
    \label{fig:strip_kde_o2o}
  \end{subfigure}
  \hfill
  \begin{subfigure}[htbp]{.155\textwidth}
    \centering
    \includegraphics[width=\linewidth]{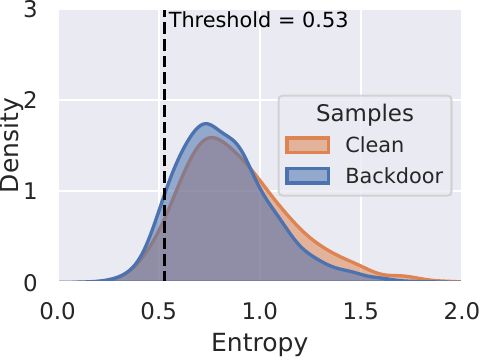}
    \caption{B2B w/ STRIP}
    \label{fig:strip_kde_b2b}
  \end{subfigure}

  \begin{subfigure}[htbp]{.155\textwidth}
    \centering
    \includegraphics[width=\linewidth]{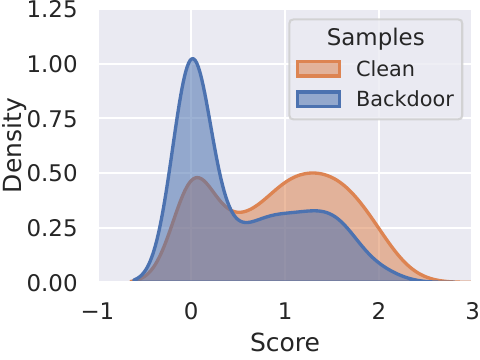}
    \caption{IBA w/ TeCo}
    \label{fig:teco_kde_iba}
  \end{subfigure}
  \hfill
  \begin{subfigure}[htbp]{.155\textwidth}
    \centering
    \includegraphics[width=\linewidth]{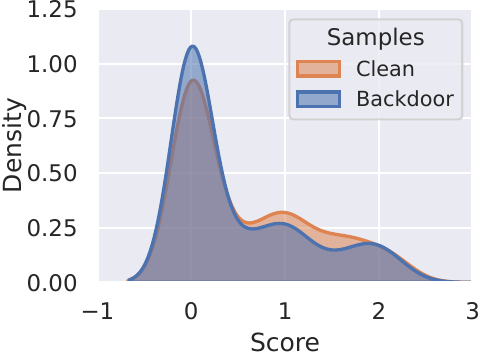}
    \caption{O2O w/ TeCo}
    \label{fig:teco_kde_o2o}
  \end{subfigure}
  \hfill
  \begin{subfigure}[htbp]{.155\textwidth}
    \centering
    \includegraphics[width=\linewidth]{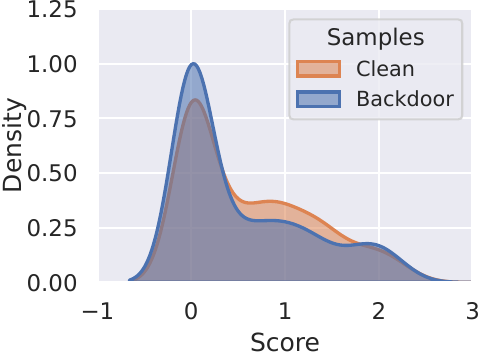}
    \caption{B2B w/ TeCo}
    \label{fig:teco_kde_b2b}
  \end{subfigure}
  \caption{KDE plots of training losses for ABL, entropy scores for STRIP, and detection scores for TeCo.}
  \label{fig:abl_strip_teco}
\end{figure}

\noindent \textbf{ABL}\cite{liAntiBackdoorLearningTraining2021} aims to separate poisoned samples from clean ones using sample training losses, and then retrain a model with the remaining data.
The process is applied progressively, weakening the spurious correlation between the trigger and the target label.
\Cref{tab:abl} reports the ASR after ABL.
The results show that ABL can mitigate simpler attacks such as HBA (reducing ASR to 15.89\%).
However, it is ineffective against the other attack variants.
In particular, for proposed attacks (e.g., B2B and B2O), ASR remains alarmingly high (above 89\%).
Moreover, the low PBA and CBA scores (0.50--0.58) suggest a notable degradation in clean-model utility.
We further explore the loss landscape and plot Kernel Density Estimation (KDE) curves in~\Cref{fig:abl_kde_iba,fig:abl_kde_o2o,fig:abl_kde_b2b}.
The results show extensive distribution overlaps between clean and poisoned samples, indicating that ABL's loss-based detection is insufficient to identify poisoned samples under our attacks reliably.

\noindent \textbf{STRIP}~\cite{gaoSTRIPDefenceTrojan2019} detects poisoned samples by measuring model prediction consistency under input perturbations.
The triggered samples are expected to maintain stable (low-entropy) outputs, while clean samples exhibit more variable (high-entropy) predictions.
To enable STRIP for semantic segmentation, we first compute per-pixel entropy maps and then aggregate them into an image-level score for detection.
\Cref{tab:strip} presents the detection results.
Overall, STRIP shows ineffective detection results.
Across all attacks, the AUC scores hover around 0.5 (ranging from 0.42 to 0.51), comparable to random guessing.
Accordingly, Recall rates are negligible, remaining consistently below 11\%, including O2O, which achieves the highest recall among the evaluated cases.
As shown in the KDE plot~\Cref{fig:strip_kde_iba,fig:strip_kde_o2o,fig:strip_kde_b2b}, the score distributions of clean and poisoned samples are nearly same.
This suggests that, in semantic segmentation, the perturbation-based signal in STRIP becomes too weak to distinguish between clean and poisoned samples.

\begin{table}[t]
  \scriptsize
  \centering
  \caption{Backdoor detection results with TeCo. $n$-std indicates different detection thresholds.}
    \begin{tabular}{c|cc|cc|cc}
    \toprule
    \multicolumn{1}{c|}{\multirow[c]{2}{*}{\textbf{Attack}}} & \multicolumn{2}{c|}{\textbf{1-std}} & \multicolumn{2}{c|}{\textbf{2-std}} & \multicolumn{2}{c}{\textbf{3-std}} \\
\cmidrule{2-7}          & \textbf{Recall} & \textbf{AUC} & \textbf{Recall} & \textbf{AUC} & \textbf{Recall} & \textbf{AUC} \\
    \midrule
    HBA     & 0.2521 & 0.5446 & 0.0652 & 0.5169 & 0.0000     & 0.5000 \\
    OFBA    & 0.2369 & 0.5574 & 0.0630 & 0.5074 & 0.0000     & 0.5000 \\
    IBA     & 0.3112 & 0.5765 & 0.0311 & 0.5117 & 0.0000     & 0.5000 \\
    O2O     & 0.1739 & 0.5018 & 0.0478 & 0.5026 & 0.0000     & 0.5000 \\
    O2B     & 0.3826 & 0.6098 & 0.0760 & 0.5242 & 0.0000     & 0.5000 \\
    B2O     & 0.1183 & 0.4829 & 0.0377 & 0.4965 & 0.0000     & 0.5000 \\
    B2B     & 0.1445 & 0.4885 & 0.0271 & 0.4971 & 0.0000     & 0.5000 \\
    INS     & 0.2134 & 0.5312 & 0.0589 & 0.5098 & 0.0000     & 0.5000 \\
    CON     & 0.2687 & 0.5523 & 0.0698 & 0.5187 & 0.0000     & 0.5000 \\
    \bottomrule
    \end{tabular}%
  \label{tab:teco}%
\end{table}%

\noindent \textbf{TeCo}~\cite{liuDetectingBackdoorsInference2023} identifies poisoned samples based on the assumption that clean images exhibit consistent robustness under diverse image corruptions.
To enable TeCo for segmentation, we apply a set of 15 image corruption operations to each input and calculate the mIoU for the clean image and its corrupted variant.
A sample is flagged as suspicious if its mIoU drop exceeds a predefined threshold.
\Cref{tab:teco} reports the detection results under thresholds of $\mu + n \cdot \sigma$, where $\mu$ and $\sigma$ denote the mean and standard deviation of the scores, respectively.
Overall, TeCo is ineffective for detecting segmentation backdoors.
Even with the most sensitive setting (1-std), AUC stays close to 0.5 (i.e., random guessing), and Recall is typically below 0.4.
To further investigate the results, we plot the KDE distributions of detection scores in~\Cref{fig:teco_kde_iba,fig:teco_kde_o2o,fig:teco_kde_b2b}.
Consistent with ABL and STRIP, the heavy overlap between clean and poisoned distributions suggests that TeCo cannot reliably distinguish between backdoored and clean samples.

\begin{table}[t]
\scriptsize
  \centering
  \caption{Backdoor detection results with Beatrix.}
      \resizebox{\linewidth}{!}{
      \begin{tabular}{c|cccc|cccc}
    \toprule
    \multicolumn{1}{c|}{\multirow[c]{2}{*}{\textbf{Attack}}} & \multicolumn{4}{c|}{\textbf{Main Class (e.g., road)}} & \multicolumn{4}{c}{\textbf{Selected Class (car)}} \\
\cmidrule{2-9}          & \textbf{ACC} & \textbf{Recall} & \textbf{F1} & \textbf{AUC} & \textbf{ACC} & \textbf{Recall} & \textbf{F1} & \textbf{AUC} \\
\midrule
    HBA   & 0.500 & 0.00  & 0.0000 & 0.500 & 0.500 & 0.01  & 0.0196 & 0.500 \\
    OFBA  & 0.500 & 0.06  & 0.1071 & 0.500 & 0.505 & 0.01  & 0.0198 & 0.505 \\
    IBA   & 0.515 & 0.10  & 0.1709 & 0.515 & 0.530 & 0.20  & 0.2985 & 0.530 \\
    O2O   & 0.525 & 0.08  & 0.1441 & 0.525 & 0.500 & 0.00  & 0.0000 & 0.500 \\
    O2B   & 0.505 & 0.01  & 0.0198 & 0.505 & 0.500 & 0.00  & 0.0000 & 0.500 \\
    B2O   & 0.515 & 0.04  & 0.0762 & 0.515 & 0.515 & 0.11  & 0.1849 & 0.515 \\
    B2B   & 0.510 & 0.06  & 0.1091 & 0.510 & 0.655 & 0.55  & 0.6145 & 0.655 \\
    INS   & 0.518 & 0.07  & 0.1251 & 0.518 & 0.510 & 0.06  & 0.1091 & 0.510 \\
    CON   & 0.512 & 0.05  & 0.0943 & 0.512 & 0.520 & 0.12  & 0.1978 & 0.520 \\
    \bottomrule
    \end{tabular}%
    }
  \label{tab:beatrix}%
\end{table}%

\noindent \textbf{Beatrix}~\cite{maBeatrixResurrectionsRobust2023} detects poisoned samples by measuring Gramian statistics of internal feature maps.
They assume that triggers can induce statistically abnormal scores compared to clean inputs.
To enable Beatrix for segmentation, we assign each image a \textit{main class}: the class that occupies the largest pixel area in the image.
We then compute the score over feature maps for the class and apply Beatrix to identify poisoned samples.
As shown in~\Cref{tab:beatrix}, this scheme results in near-random performance, with Recall approaching zero and AUC staying around 0.5 across all attacks.
We further evaluate with a \textit{selected class} (e.g., \emph{car}) to calculate the detection score.
This leads to only minor gains for most attacks, but improves B2B more noticeably.
We attribute the improvement to its background-to-background setting, which dominates the pixel distribution in images.
Therefore, the attack induces a stronger shift in global feature statistics, making Gramian statistics slightly more separable.
However, even in this case, Recall only reaches 0.55, suggesting that Beatrix remains unreliable for detecting poisoned samples in segmentation.

\noindent \textbf{Summary.}
We benchmark both existing and our proposed segmentation backdoor attacks against six representative defenses and find that:
(1) Backdoored segmentation models are often resistant to post-training defenses (e.g., fine-tuning). This robustness likely stems from the feature entanglement in dense prediction, which allows backdoor signals to persist despite parameter perturbations.
(2) Existing poisoned sample detection methods prove ineffective for segmentation. This is because segmentation attacks typically modify labels locally at the pixel level rather than global label flipping; they do not induce global statistical anomalies significant enough for reliable detection.
The effective mitigation of segmentation backdoors remains largely unresolved.

\begin{table}[t]
\scriptsize
  \centering
  \caption{Attack results with Transformers on BDD100K.}
  \resizebox{\linewidth}{!}{
    \begin{tabular}{c|c|ccc||c|c|ccc}
    \toprule
     & \textbf{Model} & \textbf{ASR $\uparrow$} & \textbf{PBA $\uparrow$} & \textbf{CBA $\uparrow$} &  & \textbf{Model} & \textbf{ASR $\uparrow$} & \textbf{PBA $\uparrow$} & \textbf{CBA $\uparrow$} \\
    \midrule
          & ViT-B  & 0.9189 & 0.6123 & 0.6287 &       & ViT-B  & 0.9187 & 0.6234 & 0.6312 \\
    O2O   & DeiT-S & 0.9145 & 0.6051 & 0.6198 &  B2B  & DeiT-S & 0.9023 & 0.6198 & 0.6245 \\
          & Swin-T & 0.9378 & 0.6115 & 0.6241 &       & Swin-T & 0.9334 & 0.6295 & 0.6267 \\
     \midrule
          & ViT-B  & 0.9247 & 0.6245 & 0.6334 &       & ViT-B  & 0.8934 & 0.6378 & 0.6489 \\
    O2B   & DeiT-S & 0.9039 & 0.6198 & 0.6256 & INS   & DeiT-S & 0.8867 & 0.6345 & 0.6412 \\
          & Swin-T & 0.9486 & 0.6232 & 0.6239 &       & Swin-T & 0.9098 & 0.6434 & 0.6441 \\
     \midrule
          & ViT-B  & 0.9712 & 0.6189 & 0.6298 &       & ViT-B  & 0.9067 & 0.6312 & 0.6378 \\
    B2O   & DeiT-S & 0.9678 & 0.6134 & 0.6221 & CON   & DeiT-S & 0.9012 & 0.6267 & 0.6289 \\
          & Swin-T & 0.9823 & 0.6172 & 0.6243 &       & Swin-T & 0.9221 & 0.6298 & 0.6323 \\
    \bottomrule
    \end{tabular}%
    }
  \label{tab:exp_transformers}%
\end{table}%

\section{Attacking Emerging Architectures}
\label{sec:attack_transformer_sam}

To answer \textbf{RQ3}, we extend our evaluation to recent segmentation architectures, including Transformer-based architectures and the Segment Anything Model (SAM).

\subsection*{Attacking Transformer-Based Models}

We attack Transformers with the experimental settings detailed in~\Cref{sec:exp_setup}, including the optimized trigger parameters and selected victim--target pairs.
\Cref{tab:exp_transformers} summarizes the results on BDD100K for ViT-B~\cite{dosovitskiyImageWorth16x162021}, DeiT-S~\cite{touvronTrainingDataefficientImage2021}, and Swin-T~\cite{liuSwinTransformerHierarchical2021} (all strucutred with a UPerNet head).
Across all settings, these architectures remain vulnerable to backdoor attacks.
Specifically, Swin-T achieves attack performance comparable to ConvNeXt-T, as they share similar feature extraction architectures.
ViT-B and DeiT exhibit slightly lower ASRs, but still remain clearly susceptible to backdoor attacks.
Meanwhile, PBAs and CBAs consistently stay above 0.6, indicating that the attacks preserve the model utility on clean inputs.
More details are presented in~\Cref{sec:transformers_additional}.

\begin{table}[t]
\scriptsize
  \centering
  \caption{Attack results for SAM~\cite{raviSAM2Segment2024}.}
    \begin{tabular}{c|ccc}
    \toprule
    \textbf{Attack} & Mask-Distortion & Mask-Erasure & Mask-Injection \\
    \midrule
    \textbf{ASR} & 0.9134 & 0.9082 & 0.9166 \\
    \bottomrule
    \end{tabular}%
  \label{tab:sam_attack}%
\end{table}%

\subsection*{Attacking Segment Anything Model (SAM)}

\noindent \textbf{Motivation.}
The Segment Anything Model (SAM)~\cite{kirillovSegmentAnything2023,raviSAM2Segment2024} introduces a promptable framework for image segmentation.
This enables users to generate high-quality masks with simple inputs, such as points or bounding boxes.
Trained on massive-scale datasets, SAM is capable of zero-shot segmentation on unseen images without additional fine-tuning.

Unlike conventional segmentation models, SAM adopts a distinct architecture and is trained on large-scale datasets.
Instead of predicting a fixed set of semantic labels, SAM learns to segment objects in a category-agnostic manner: given a prompt, it predicts the corresponding object mask without assigning a semantic class label.
This promptable interface enables flexible interaction, allowing users to segment arbitrary objects as long as they are visually distinguishable in the image.

\noindent \textbf{Proposed Attacks.}
To deploy backdoor attacks against SAM, we adapt BADSEG to match SAM's prompt-conditioned mask prediction.
Instead of inducing label misclassification, our proposed attacks directly manipulate the predicted masks.
Specifically, we consider three attack vectors tailored to SAM: (1) \textit{Mask-Distortion}, where the trigger alters the shape, location, or extent of the predicted mask; (2) \textit{Mask-Erasure}, where the trigger erases the output, causing the mask to vanish entirely; and (3) \textit{Mask-Injection}, where the trigger fabricates spurious masks in regions where no object exists.

For the detailed attack procedure, although BADSEG is formulated for label misclassification, its trigger optimization can be adapted to mask manipulation in SAM.
Therefore, we propose BADSEG-SAM, which uses the first two stages of BADSEG to optimize trigger parameters for mask manipulation.
For each attack vector, BADSEG-SAM optimizes the trigger parameters utilizing a surrogate model.
And then it modifies the target masks to match the objective of each attack vector.
This approach allows us to effectively launch all three proposed attack vectors on SAM.

\noindent \textbf{Attack Results.}
We evaluate the robustness of SAM under three proposed attacks on the LabPicsV1 dataset~\cite{eppelComputerVisionRecognition2020}.
\Cref{tab:sam_attack} shows the results.
Overall, SAM is highly vulnerable: all three attack vectors yield high ASRs, indicating that backdoor triggers can consistently manipulate prompt-conditioned mask predictions.
Notably, even the weakest case (\textit{Mask-Erasure}) attains an ASR of 0.9082, indicating that large-scale segmentation models remain vulnerable to our attacks.

\section{Related Work}

In this section, we first review backdoor attacks in general, summarize existing work on backdoors in semantic segmentation, and then introduce backdoor defense strategies.

\noindent \textbf{Backdoor Attacks.}
Backdoor attacks were first introduced in image classification~\cite{guBadNetsIdentifyingVulnerabilities2019}, where training samples were manipulated by injecting a patch trigger and relabeling them to a target class.
Subsequent work diversified trigger design, such as cartoon blending~\cite{chenTargetedBackdoorAttacks2017}, invisible perturbations~\cite{zhongBackdoorEmbeddingConvolutional2020}, and reflections~\cite{liuReflectionBackdoorNatural2020}.
Beyond image classification, backdoor attacks have also been explored in diverse settings, such as federated learning~\cite{wangAttackTailsYes2020}, natural language processing~\cite{chenBadNLBackdoorAttacks2021}, and object detection~\cite{chanBadDetBackdoorAttacks2022}.

\noindent \textbf{Backdoor Attacks on Semantic Segmentation.}
Several studies have demonstrated the feasibility of backdoor attacks in semantic segmentation with different trigger designs, such as black line triggers in Hidden Backdoor Attacks (HBA)~\cite{liHiddenBackdoorAttack2021},  grid-shaped triggers in Object-Free Backdoor Attacks (OFBA)~\cite{maoObjectfreeBackdoorAttack2023}, and a ``Hello-Kitty'' logos in Influencer Backdoor Attacks (IBA)~\cite{lanInfluencerBackdoorAttack2023}.
While these works establish initial segmentation backdoor studies, they are restricted to narrow settings.
More recently, research has explored adversarial attacks in the Segment Anything Model (SAM)~\cite{zhouDarkSAMFoolingSegment2024}.
In contrast to these existing studies, we revisit segmentation backdoor attacks through a broader lens, examining potential threats, trigger designs, defense benchmarking, and emerging architectures.

\noindent \textbf{Backdoor Defenses.}
Various defenses have been proposed to mitigate backdoor attacks.
One line of work applies post-training interventions to suppress backdoor behavior~\cite{liuNeuralTrojans2017,liAntiBackdoorLearningTraining2021,chenEffectiveBackdoorDefense2022,shaFineTuningAllYou2022,gaoBackdoorDefenseAdaptively2023}.
Another line focuses on detecting or neutralizing backdoors in already trained models~\cite{gaoSTRIPDefenceTrojan2019,zhengDataFreeBackdoorRemoval2022,liNeuralAttentionDistillation2021,liuDetectingBackdoorsInference2023,maBeatrixResurrectionsRobust2023}.
While these approaches have shown strong results for image classification, their effectiveness for semantic segmentation remains largely unexplored.
In this work, we benchmark existing and the proposed attacks against six representative defenses, and find that they offer limited protection for segmentation backdoor attacks. Additional discussion is presented in~\Cref{sec:related_work_additional}.

\section{Conclusion}

This work presents the first comprehensive study of backdoor attacks on semantic segmentation.
We revisit the threats and identify four coarse-grained attacks and two fine-grained attacks, exposing previously overlooked backdoor vulnerabilities.
To launch these attacks, we propose BADSEG, a unified framework that integrates trigger parameter optimization and label manipulation strategies.
Extensive experiments across multiple architectures and datasets demonstrate that BADSEG achieves consistently high attack success rates while preserving model utility.
Furthermore, we further benchmark existing attacks against six representative defenses, revealing that these defenses provide limited protection against the proposed attacks.
Finally, we show that these vulnerabilities persist in recent emerging architectures, including transformers and SAM.
Our results expose overlooked threats in segmentation, motivate the development of efficient backdoor defenses, and establish a practical benchmark for developing more secure segmentation models.

{
\bibliographystyle{plain}
\bibliography{selected.bib}
}

\appendix

\section{Additional Preliminaries}
\label{sec:additional_pre}

\noindent \textbf{Semantic Segmentation Pipeline.}
Semantic segmentation can be used in many applications, such as autonomous driving~\cite{siamComparativeStudyRealTime2018,fengDeepMultiModalObject2021,guCLFTCameraLiDARFusion2024}, medical imaging~\cite{bougourziRecentAdvancesMedical2025,wieczorekTransformerBasedSemantic2025}, and remote sensing~\cite{wangSAMRSScalingupRemote2023}.
We take autonomous driving as a representative application scenario to explain the semantic segmentation pipeline.
In autonomous driving, semantic segmentation enables the perception of complex urban environments by identifying critical elements, such as drivable surfaces, pedestrians, and vehicles, from camera images.
The pipeline involves acquiring data from public datasets with pixel-level annotations, followed by preprocessing and augmentation to enhance robustness.
A deep learning model is trained using cross-entropy loss to predict segmentation maps that assign semantic labels to every pixel.
These maps integrate into the vehicle's perception and planning systems, enabling safe navigation by detecting drivable areas and obstacles.
Attackers can exploit this pipeline by injecting backdoor triggers during data acquisition, creating backdoored models that compromise the vehicle's perception and planning capabilities.

\begin{table}[t]
\scriptsize
  \centering
  \caption{List of candidate options for trigger parameters.}
    \begin{tabular}{c|cl}
    \toprule
    \textbf{Attack} & \textbf{Attribute} & \textbf{Options} \\
    \midrule
    \multirow{5}[2]{*}{O2O} & Shape & circle, square, triangle, batman logo \\
          & Size  & 1/12, 1/10, 1/8, 1/6, 1/4, 1/2 \\
          & Position & object center, random on object, random outside object \\
          & Quantity & 1, 2, 3, 4, 5 \\
          & Intensity & 0.2, 0.3, 0.4, 0.5, 0.6, 0.7, 0.8 \\
    \midrule
    \multirow{5}[2]{*}{O2B} & Shape & circle, square, triangle, batman logo \\
          & Size  & 1/12, 1/10, 1/8, 1/6, 1/4, 1/2 \\
          & Position & object center, random on object, random outside object \\
          & Quantity & 1, 2, 3, 4, 5 \\
          & Intensity & 0.2, 0.3, 0.4, 0.5, 0.6, 0.7, 0.8 \\
    \midrule
    \multirow{5}[2]{*}{B2O} & Shape & circle, square, triangle, batman logo \\
          & Size  & 1/12, 1/10, 1/8, 1/6, 1/4, 1/2 \\
          & Position & - \\
          & Quantity & 1, 2, 3, 4, 5 \\
          & Intensity & 0.2, 0.3, 0.4, 0.5, 0.6, 0.7, 0.8 \\
    \midrule
    \multirow{5}[2]{*}{B2B} & Shape & circle, square, triangle, batman logo \\
          & Size  & 0.005, 0.010, 0.015, 0.020, 0.025 \\
          & Position & - \\
          & Quantity & 1, 3, 5, 7, 10 \\
          & Intensity & 0.2, 0.3, 0.4, 0.5, 0.6, 0.7, 0.8 \\
    \bottomrule
    \end{tabular}%
  \label{tab:trigger_candidate_list}%
\end{table}%

\section{Additional Details in BADSEG}
\label{sec:badseg_additional}

\noindent \textbf{Discussion on candidate trigger options.}
\Cref{tab:trigger_candidate_list} summarizes the discrete search space for trigger parameters in Stage 2.
Our candidates are selected to strike a balance between attack effectiveness and stealthiness.
The search spaces are shared across attack vectors, with minor adjustments when the threat changes.
For background-oriented attacks (B2O/B2B), the trigger is positioned relative to the specified background/victim region; thus, the object-centric \emph{Position} options used in O2O/O2B are not applicable (marked as ``--'').

\noindent \textbf{Details for optimization of discrete parameters.}
We employ the \emph{Gumbel-Softmax} reparameterization~\cite{jangCategoricalReparameterizationGumbelSoftmax2017,maddisonConcreteDistributionContinuous2017}.
We maintain a categorical distribution for each discrete parameter $\lambda_p$.
Specifically, let $\mathcal{V}_p=\{v^{(p)}_i\}_{i=1}^{m_p}$ denote its candidate set and let
$\boldsymbol{\xi}^{(p)}=(\xi^{(p)}_1,\ldots,\xi^{(p)}_{m_p})$ be the associated categorical probabilities.
Using the Gumbel--Softmax reparameterization, we draw i.i.d.\ Gumbel noise $\mathcal{G}^{(p)}_i$ and obtain a differentiable, near one-hot sample $\eta^{(p)}$ shown in~\Cref{eq:gumbel}.

We then map $\eta^{(p)}$ to a relaxed parameter value via a weighted combination of candidates:
\begin{equation}
\tilde{\lambda}_p=\sum_{i=1}^{m_p}\eta^{(p)}_i\, v^{(p)}_i,
\qquad
\tilde{\boldsymbol{\lambda}}=(\tilde{\lambda}_1,\ldots,\tilde{\lambda}_k).
\label{eq:relax_additional}
\end{equation}
Constructing $\delta=\mathcal{F}(\tilde{\boldsymbol{\lambda}})$ makes the trigger differentiable with respect to $\{\boldsymbol{\xi}^{(p)}\}_{p=1}^k$.
Accordingly, we optimize:
\begin{equation}
\min_{\{\boldsymbol{\xi}^{(p)}\}_{p=1}^k}\;
\mathbb{E}_{\{\mathcal{G}^{(p)}\}}
\left[
\sum_{(x, y)\in\mathbf{D}_s}
\mathcal{L}\Big(
\mathbf{S}\big(\mathcal{T}(x,\mathcal{F}(\tilde{\boldsymbol{\lambda}}))\big),\, y^t
\Big)
\right].
\label{eq:all_additional}
\end{equation}
$\tilde{\boldsymbol{\lambda}}$ is computed from $\{\boldsymbol{\xi}^{(p)}\}$, so the objective is differentiable w.r.t. $\{\boldsymbol{\xi}^{(p)}\}$.
$\mathbb{E}_{\{\mathcal{G}^{(p)}\}}$ denotes the expectation over the random Gumbel noises used to sample $\boldsymbol{\eta}^{(p)}$ for each discrete trigger parameter in the Gumbel--Softmax relaxation (approximated in practice by drawing one sample per iteration).

After optimization, we discretize each parameter by selecting the most likely option
$\hat{\lambda}_p = v^{(p)}_{\arg\max_i \xi^{(p)}_i}$ and construct the final trigger
$\delta=\mathcal{F}(\hat{\boldsymbol{\lambda}})$.

\section{Additional Experimental Setup}
\label{sec:exp_setup_additional}

\noindent \textbf{Datasets.}
We evaluate BADSEG on two widely used autonomous driving benchmarks.
BDD100K~\cite{yuBDD100KDiverseDriving2020} contains 100,000 images under diverse driving conditions (time of day, weather, and scene types), of which 10,000 are annotated for semantic segmentation.
Cityscapes~\cite{cordtsCityscapesDatasetSemantic2016} consists of 5,000 finely annotated images collected from 50 cities, with 2,975 for training, 500 for validation, and 1,525 for testing.
We primarily conduct our experiments on BDD100K, as it is the more complex dataset with diverse driving scenes and illumination conditions.
Evaluating BADSEG on BDD100K enables us to assess its robustness under challenging real-world scenarios.
Results on Cityscapes are also reported to verify that our findings transfer across datasets with different scales and collection environments.

In~\Cref{sec:attack_transformer_sam}, we evaluate BADSEG-SAM on LabPicsV1~\cite{eppelComputerVisionRecognition2020}.
Vector-LabPics V1 contains 2,187 images of chemical experiments featuring materials in mostly transparent vessels across diverse laboratory scenes and everyday conditions (e.g., beverage handling).
We choose this dataset for SAM evaluation because it provides region-level annotations for each material phase along with its type, enabling us to assess SAM’s ability to produce fine-grained, detailed segmentations.

\noindent \textbf{Models.}
We consider three representative semantic segmentation architectures.
PSPNet~\cite{zhaoPyramidSceneParsing2017} employs a pyramid pooling module to aggregate multi-scale contextual information, standing as one of the earliest baselines in semantic segmentation.
DeepLabV3~\cite{chenEncoderDecoderAtrousSeparable2018} integrates atrous spatial pyramid pooling for enhanced contextual reasoning.
Both utilize ResNet backbones that are pre-trained on ImageNet.
ConvNeXt-T~\cite{liuConvNet2020s2022}, pre-trained on ImageNet-22K, modernizes ConvNet design by incorporating architectural refinements inspired by transformers while retaining convolutional efficiency.
We conduct most experiments using ConvNeXt-T, for it is a modern architecture with high efficiency, making it a strong baseline for semantic segmentation.

In~\Cref{sec:attack_transformer_sam}, we consider three Transformer-based emerging architectures, including ViT-B~\cite{dosovitskiyImageWorth16x162021}, DeiT-S~\cite{touvronTrainingDataefficientImage2021}, and Swin-T~\cite{liuSwinTransformerHierarchical2021} (all strucutred with a UPerNet head).
ViT-B~\cite{dosovitskiyImageWorth16x162021} is a Vision Transformer that tokenizes an image into fixed-size patches and models global context via full self-attention, offering strong representation capacity.
DeiT-S~\cite{touvronTrainingDataefficientImage2021} follows the same transformer formulation as ViT but improves data efficiency through knowledge distillation during training, making it a lightweight yet competitive alternative under limited data.
Swin-T~\cite{liuSwinTransformerHierarchical2021} adopts a hierarchical design with window-based self-attention and shifted windows, yielding multi-scale features that better match the locality and scale variation required by segmentation while keeping attention computation tractable.
These transformer-based models are widely adopted in prior work and thus serve as representative baselines in our evaluation.

We also consider emerging architectures such as SAM.
Specifically, we use sam2-hiera-T~\cite{raviSAM2Segment2024} in our evaluation. sam2-hiera-T is a lightweight SAM 2 variant that adopts a hierarchical transformer backbone (``Hiera''), enabling strong mask prediction with improved efficiency and making it a practical representative of recent segment-anything–style models.

\noindent \textbf{Evaluation Metrics.}
Following prior backdoor evaluations in semantic segmentation~\cite{maoObjectfreeBackdoorAttack2023,lanInfluencerBackdoorAttack2023}, we assess attacks in the following metrics:
\begin{itemize}[noitemsep, topsep=0pt]

\item \textit{Attack Success Rate (ASR).}
ASR measures how often victim pixels are successfully mis-segmented to the specified target class under the poisoned test:
$\text{ASR} = {N_\text{success}} / {N_\text{victim}}$,
where $N_{\text{victim}}$ is the number of victim pixels and $N_{\text{success}}$ is the subset predicted as the target class.
$N_{\text{victim}}$ is attack-dependent: for \emph{B2O} it is restricted to the intended appearing region; for \emph{Instance-Level} attacks it is restricted to the targeted instance; otherwise it includes all pixels of the victim class.

\item  \textit{Poisoned Benign Accuracy (PBA).}
PBA captures segmentation utility on \emph{non-victim} pixels in the poisoned test.
We compute it as mIoU between predictions and ground truth after masking out victim pixels.

\item  \textit{Clean Benign Accuracy (CBA).}
CBA measures standard segmentation utility on clean data, defined as the mIoU on the unmodified test set.
An effective backdoor should keep CBA close to a cleanly trained model, indicating minimal impact on normal functionality.
\end{itemize}

We measure attack stealthiness using three standard image-similarity metrics computed between clean and poisoned images.
\begin{itemize}[noitemsep, topsep=0pt]
\item  \textit{Peak Signal-to-Noise Ratio (PSNR)} summarizes pixel-level distortion, where higher values indicate smaller perturbations.

\item  \textit{Structural Similarity Index(SSIM)} measures similarity in luminance, contrast, and structure, with higher scores indicating better perceptual alignment.

\item  \textit{Learned Perceptual Image Patch Similarity (LPIPS)} uses deep features to approximate human perceptual distance; lower LPIPS means the poisoned image is more perceptually similar to the clean one.
\end{itemize}

\section{Additional Experiments}
\label{sec:exp_additional}

\begin{table}[t]
\centering
\scriptsize
\caption{Top 20 closest class pairs by normalized distance with different models (complementary to~\Cref{tab:ablation_class_pair_simi}).}
\label{tab:ablation_top20_class_pair_additional}

\begin{subtable}[t]{\linewidth}

\scriptsize
  \centering
  \caption{Results with DeepLabV3.}
    \begin{tabular}{cc|cc}
    \toprule
    \textbf{Rank} & \textbf{Class Pair} & \textbf{Distance} & \textbf{Suitable Attacks} \\
    \midrule
    1 &(Building, Vegetation) & 0.1087 & B2B \\
    2 &(Building, Traffic Sign) & 0.1187 & B2B \\
    3 &(Vegetation, Traffic Sign) & 0.1414 & B2B \\
    4 &(Car, Sidewalk) & 0.1511 & O2B, B2O \\
    5 &(Sidewalk, Road) & 0.1570 & B2B \\
    6 &(Sidewalk, Person) & 0.1580 & O2B, B2O \\
    7 &(Road, Car) & 0.1591 & O2B, B2O \\
    8 &(Sidewalk, Terrain) & 0.1760 & B2B \\
    9 &(Fence, Building) & 0.1815 & B2B \\
    10 &(Pole, Terrain) & 0.1837 & B2B \\
    11 &(Person, Car) & 0.1875 & O2O \\
    12 &(Building, Pole) & 0.2015 & B2B \\
    13 &(Traffic Sign, Fence) & 0.2100 & B2B \\
    14 &(Person, Terrain) & 0.2112 & O2B, B2O \\
    15 &(Pole, Traffic Sign) & 0.2125 & B2B \\
    16 &(Road, Person) & 0.2232 & O2B, B2O \\
    17 &(Person, Fence) & 0.2269 & O2B, B2O \\
    18 &(Vegetation, Pole) & 0.2277 & B2B \\
    19 &(Sidewalk, Pole) & 0.2407 & B2B \\
    20 &(Traffic Light, Building) & 0.2412 & B2B \\
    \bottomrule
    \end{tabular}%
  \label{tab:top20_class_pairs_dl}%
\end{subtable}%
\vfill
\begin{subtable}[t]{\linewidth}
\scriptsize
  \centering
  \caption{Results with PSPNet.}
    \begin{tabular}{cc|cc}
    \toprule
    \textbf{Rank} & \textbf{Class Pair} & \textbf{Distance} & \textbf{Suitable Attacks} \\
    \midrule
    1 &(Building, Vegetation) & 0.1116 & B2B \\
    2 &(Building, Traffic Sign) & 0.1207 & B2B \\
    3 &(Vegetation, Traffic Sign) & 0.1408 & B2B \\
    4 &(Person, Sidewalk) & 0.1513 & O2B, B2O \\
    5 &(Car, Sidewalk) & 0.1523 & O2B, B2O \\
    6 &(Car, Road) & 0.1563 & O2B, B2O \\
    7 &(Sidewalk, Road) & 0.1569 & B2B \\
    8 &(Sidewalk, Terrain) & 0.1721 & B2B \\
    9 &(Fence, Building) & 0.1758 & B2B \\
    10 &(Car, Person) & 0.1759 & O2O \\
    11 &(Pole, Terrain) & 0.1814 & B2B \\
    12 &(Pole, Building) & 0.1964 & B2B \\
    13 &(Terrain, Person) & 0.2059 & O2B, B2O \\
    14 &(Road, Person) & 0.2078 & O2B, B2O \\
    15 &(Fence, Traffic Sign) & 0.2082 & B2B \\
    16 &(Traffic Sign, Pole) & 0.2092 & B2B \\
    17 &(Fence, Person) & 0.2150 & O2B, B2O \\
    18 &(Pole, Vegetation) & 0.2240 & B2B \\
    19 &(Sidewalk, Pole) & 0.2251 & B2B \\
    20 &(Pole, Car) & 0.2268 & O2B, B2O \\
    \bottomrule
    \end{tabular}%
  \label{tab:top20_class_pairs_ps}%
\end{subtable}%
\end{table}%

\noindent \textbf{Additional Results on the Impact of Surrogate Models.}
\Cref{tab:ablation_top20_class_pair_additional} lists the top 20 closest class pairs by normalized distance for DeepLabV3 and PSPNet.
These results align closely with ConvNeXt-T, as our preferred class pairs consistently appear in the top 20 across all models.
This demonstrates that optimized class pair selection remains robust regardless of the surrogate model used.

\begin{table}[t]
\centering
\scriptsize
\caption{Attack performance under different victim--target class configurations (fixed victim/target) for various attacks(complementary to~\Cref{tab:ablation_poisoning_o2b}).}
\label{tab:additional_poisoning_three_subtables_additional}
\begin{subtable}[t]{\linewidth}
\scriptsize
\centering
\caption{O2O: fixed victim (car), varying target (objects).}
\label{tab:poisoning_o2o_victim_unchanged_additional}
\begin{tabular}{ll|ccc}
\toprule
\textbf{Victim} & \textbf{Target} & \textbf{ASR $\uparrow$} & \textbf{PBA $\uparrow$} & \textbf{CBA $\uparrow$} \\
\midrule
      & person     & \textbf{0.9352} & 0.6090 & 0.6265 \\
      & rider      & 0.9163          & 0.6102 & 0.6316 \\
      & truck      & 0.9307          & 0.6310 & 0.6369 \\
car   & bus        & 0.9184          & 0.6223 & 0.6382 \\
      & train      & 0.9281          & 0.6366 & 0.6367 \\
      & motorcycle & 0.9264          & 0.6006 & 0.6310 \\
      & bicycle    & 0.9303          & 0.5874 & 0.6227 \\
\bottomrule
\end{tabular}%
\end{subtable}
\vfill
\begin{subtable}[t]{\linewidth}
\centering
\caption{O2B: fixed victim (car), varying target (stuff).}
\label{tab:poisoning_o2b_victim_unchanged_sub_additional}
\begin{tabular}{ll|ccc}
\toprule
\textbf{Victim} & \textbf{Target} & \textbf{ASR $\uparrow$} & \textbf{PBA $\uparrow$} & \textbf{CBA $\uparrow$} \\
\midrule
       & road          & \textbf{0.9428} & 0.6283 & 0.6310 \\
       & sidewalk      & 0.9280 & 0.6214 & 0.6356 \\
       & building      & 0.9288 & 0.6197 & 0.6274 \\
       & wall          & 0.9058 & 0.6245 & 0.6195 \\
       & fence         & 0.9083 & 0.6228 & 0.6261 \\
car    & pole          & 0.9187 & 0.6206 & 0.6280 \\
       & traffic light & 0.9246 & 0.6116 & 0.6387 \\
       & traffic sign  & 0.9275 & 0.6069 & 0.6330 \\
       & vegetation    & 0.9427 & 0.6183 & 0.6147 \\
       & terrain       & 0.9154 & 0.6219 & 0.6281 \\
       & sky           & 0.9250 & 0.6374 & 0.6421 \\
\bottomrule
\end{tabular}%
\end{subtable}
\vfill
\begin{subtable}[t]{\linewidth}
\centering
\caption{O2B: fixed target (road), varying victim (objects).}
\label{tab:poisoning_o2b_target_unchanged_sub_additional}
\begin{tabular}{ll|ccc}
\toprule
\textbf{Victim} & \textbf{Target} & \textbf{ASR $\uparrow$} & \textbf{PBA $\uparrow$} & \textbf{CBA $\uparrow$} \\
\midrule
person     &         & 0.7498 & 0.6422 & 0.6075 \\
rider      &         & 0.4054 & 0.6516 & 0.6151 \\
car        &         & \textbf{0.9428} & 0.6283 & 0.6310 \\
truck      & road    & 0.7264 & 0.6315 & 0.6147 \\
bus        &         & 0.7850 & 0.6216 & 0.6163 \\
train      &         & 0.0143 & 0.6530 & 0.6185 \\
motorcycle &         & 0.4051 & 0.6239 & 0.6125 \\
bicycle    &         & 0.7968 & 0.6290 & 0.5942 \\
\bottomrule
\end{tabular}%
\end{subtable}
\end{table}

\noindent \textbf{Additional Results on the Impact of Victim--Target Classes.}
\Cref{tab:additional_poisoning_three_subtables_additional} provides additional information for attack performance under different victim/target class configurations.
\Cref{tab:poisoning_o2o_victim_unchanged_additional} evaluates O2O Attacks with a fixed victim class (car) and varying target classes.
Attack effectiveness remains consistently high, with over 91\% of poisoned samples successfully mis-segmented into target classes, while model utility remains stable.
This demonstrates that once a victim class is compromised, the backdoor generalizes effectively across multiple target classes without degrading model utility.
\Cref{tab:poisoning_o2b_victim_unchanged_sub_additional,tab:poisoning_o2b_target_unchanged_sub_additional} report additional model-utility results, showing that varying the victim or target class mainly affects backdoor activation, while leaving clean-data performance essentially unchanged.

\begin{table}[t]
\scriptsize
  \centering
  \caption{Results of various poisoning rates with ASR metrics.}
    \begin{tabular}{c|ccccc}
    \toprule
    \textbf{Rate} & 0.05  & 0.1   & 0.2   & 0.3   & 0.5 \\
    \midrule
    \textbf{O2O} & 0.8936 & 0.9212 & 0.9352 & 0.9346 & 0.9495 \\
    \textbf{O2B} & 0.8794 & 0.9257 & 0.9428 & 0.9406 & 0.9423 \\
    \textbf{B2O} & 0.9349 & 0.9621 & 0.9835 & 0.9877 & 0.9892 \\
    \textbf{B2B} & 0.8996 & 0.915 & 0.9315 & 0.9267 & 0.9364 \\
    \textbf{INS} & 0.8534 & 0.8891 & 0.9076 & 0.9143 & 0.9289 \\
    \textbf{CON} & 0.8623 & 0.8956 & 0.9204 & 0.9267 & 0.9378 \\
    \bottomrule
    \end{tabular}%
  \label{tab:exp_poisoning_rate_asr}%
\end{table}%

\noindent \textbf{Impact of Poisoning Rate.}
\Cref{tab:exp_poisoning_rate_asr} studies the effect of the poisoning rate (the percentage of training samples stamped with triggers) on attack performance.
ASR increases monotonically as the poisoning rate grows.
A similar trend is observed across all attack vectors, with B2O Attacks consistently reaching the highest ASRs.
Notably, our attacks remain highly effective even at a low poisoning rate (0.05), whereas prior work typically relied on 0.1 to 0.2 to obtain practical backdoor performance~\cite{maoObjectfreeBackdoorAttack2023,lanInfluencerBackdoorAttack2023}.
This highlights the efficiency of our poisoning strategy.

\begin{table}[t]
\scriptsize
  \centering
  \caption{Comparison between prior work and our attacks.}
  \resizebox{\linewidth}{!}{
    \begin{tabular}{c|ccccc}
    \toprule
\textbf{Method} & \textbf{HBA~\cite{liHiddenBackdoorAttack2021}} & \textbf{OFBA~\cite{maoObjectfreeBackdoorAttack2023}} & \textbf{IBA~\cite{lanInfluencerBackdoorAttack2023}} & \textbf{Ours (O2O)} & \textbf{Ours (O2B)} \\
    \midrule
    ASR   & 0.1815 & 0.7859 & 0.8314 & 0.9352 & 0.9428 \\
    \midrule
    \textbf{Method} & \textbf{Ours (B2O)} & \textbf{Ours (B2B)} & \textbf{Ours (Ins)} & \textbf{Ours (Con)} &  \\
    \midrule
    ASR   & 0.9835 & 0.9315 & 0.9076 & 0.9204 &  \\
    \bottomrule
    \end{tabular}
}
  \label{tab:exp_prior_work}%
\end{table}%

\noindent \textbf{Comparison with Prior Work.}
We compare against three representative baselines: HBA~\cite{liHiddenBackdoorAttack2021}, OFBA~\cite{maoObjectfreeBackdoorAttack2023}, and IBA~\cite{lanInfluencerBackdoorAttack2023}.
In our studies, all three methods fall under the category of O2B Attacks.
We re-implement these baselines under consistent hyperparameter settings to ensure fair comparison and evaluate them using ConvNeXt-T on BDD100K.
The results in~\Cref{tab:exp_prior_work} show that our proposed attacks consistently outperform prior approaches, confirming the effectiveness of our framework.

\section{Additional Experiments and Details on Defenses}
\label{sec:defense_additional}

\begin{figure}[t]
  \scriptsize
  \begin{subfigure}[htbp]{.155\textwidth}
    \centering
    \includegraphics[width=\linewidth]{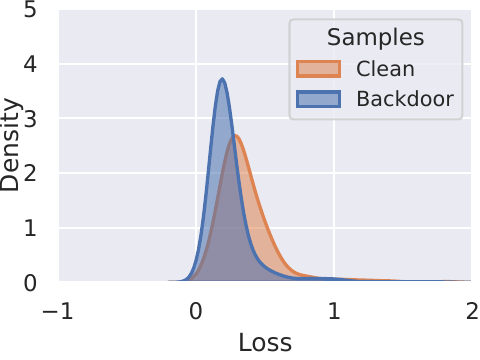}
    \caption{HBA}
    \label{fig:abl_kde_hba2}
  \end{subfigure}
  \hfill
  \begin{subfigure}[htbp]{.155\textwidth}
    \centering
    \includegraphics[width=\linewidth]{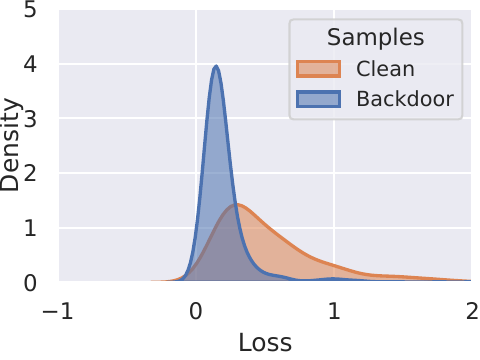}
    \caption{OFBA}
    \label{fig:abl_kde_ofba2}
  \end{subfigure}
  \hfill
  \begin{subfigure}[htbp]{.155\textwidth}
    \centering
    \includegraphics[width=\linewidth]{figures/fig_abl_kde_iba.pdf}
    \caption{IBA}
    \label{fig:abl_kde_iba2}
  \end{subfigure}
  \vfill
  \begin{subfigure}[htbp]{.155\textwidth}
    \centering
    \includegraphics[width=\linewidth]{figures/fig_abl_kde_o2o.pdf}
    \caption{O2O}
    \label{fig:abl_kde_o2o2}
  \end{subfigure}
  \hfill
  \begin{subfigure}[htbp]{.155\textwidth}
    \centering
    \includegraphics[width=\linewidth]{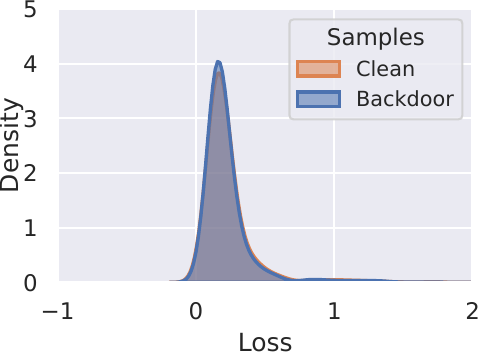}
    \caption{O2B}
    \label{fig:abl_kde_o2b2}
  \end{subfigure}
  \hfill
  \begin{subfigure}[htbp]{.155\textwidth}
    \centering
    \includegraphics[width=\linewidth]{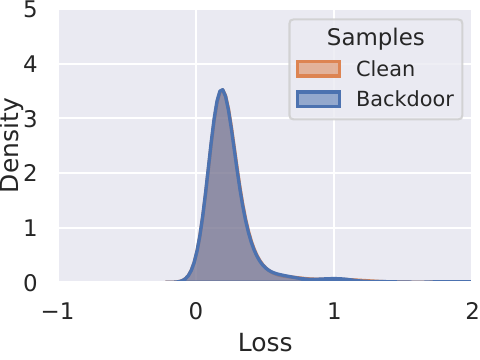}
    \caption{B2O}
    \label{fig:abl_kde_b2o2}
  \end{subfigure}
  \vfill
  \begin{subfigure}[htbp]{.155\textwidth}
    \centering
    \includegraphics[width=\linewidth]{figures/fig_abl_kde_b2b.pdf}
    \caption{B2B}
    \label{fig:abl_kde_b2b2}
  \end{subfigure}
  \hfill
  \begin{subfigure}[htbp]{.155\textwidth}
    \centering
    \includegraphics[width=\linewidth]{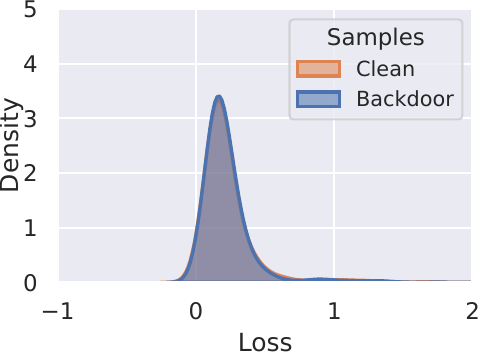}
    \caption{INS}
    \label{fig:abl_kde_ins2}
  \end{subfigure}
  \hfill
  \begin{subfigure}[htbp]{.155\textwidth}
    \centering
    \includegraphics[width=\linewidth]{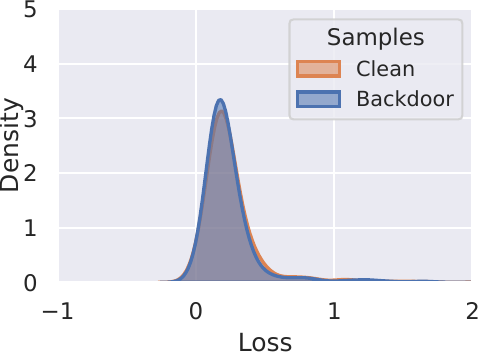}
    \caption{CON}
    \label{fig:abl_kde_con2}
  \end{subfigure}
  \caption{KDE plots of the losses of more attacks under ABL defense (complementary to~\Cref{fig:abl_strip_teco}).}
  \label{fig:abl_additional}
\end{figure}

\begin{figure}[t]
  \scriptsize
  \begin{subfigure}[htbp]{.155\textwidth}
    \centering
    \includegraphics[width=\linewidth]{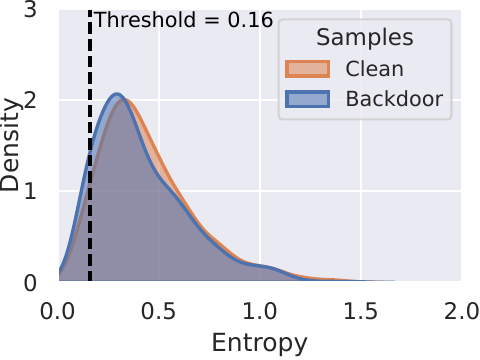}
    \caption{HBA}
    \label{fig:strip_kde_hba2}
  \end{subfigure}
  \hfill
  \begin{subfigure}[htbp]{.155\textwidth}
    \centering
    \includegraphics[width=\linewidth]{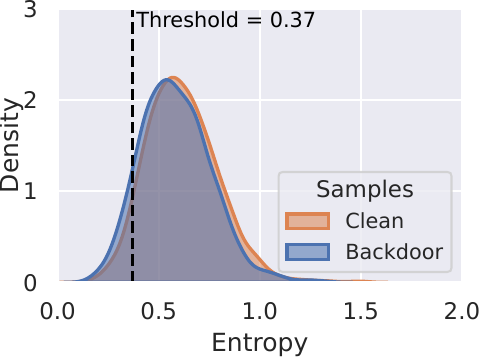}
    \caption{OFBA}
    \label{fig:strip_kde_ofba2}
  \end{subfigure}
  \hfill
  \begin{subfigure}[htbp]{.155\textwidth}
    \centering
    \includegraphics[width=\linewidth]{figures/fig_strip_kde_iba.pdf}
    \caption{IBA}
    \label{fig:strip_kde_iba2}
  \end{subfigure}
  \vfill
  \begin{subfigure}[htbp]{.155\textwidth}
    \centering
    \includegraphics[width=\linewidth]{figures/fig_strip_kde_o2o.pdf}
    \caption{O2O}
    \label{fig:strip_kde_o2o2}
  \end{subfigure}
  \hfill
  \begin{subfigure}[htbp]{.155\textwidth}
    \centering
    \includegraphics[width=\linewidth]{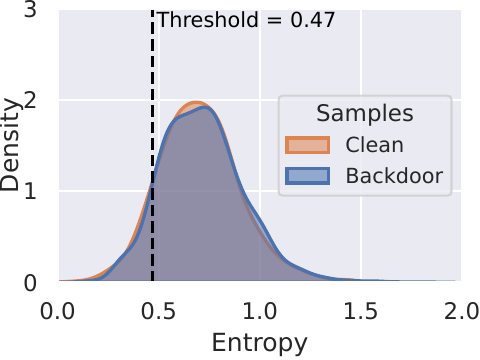}
    \caption{O2B}
    \label{fig:strip_kde_o2b2}
  \end{subfigure}
  \hfill
  \begin{subfigure}[htbp]{.155\textwidth}
    \centering
    \includegraphics[width=\linewidth]{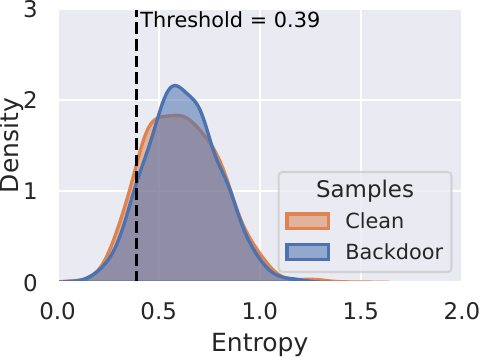}
    \caption{B2O}
    \label{fig:strip_kde_b2o2}
  \end{subfigure}
  \vfill
  \begin{subfigure}[htbp]{.155\textwidth}
    \centering
    \includegraphics[width=\linewidth]{figures/fig_strip_kde_b2b.pdf}
    \caption{B2B}
    \label{fig:strip_kde_b2b2}
  \end{subfigure}
  \hfill
  \begin{subfigure}[htbp]{.155\textwidth}
    \centering
    \includegraphics[width=\linewidth]{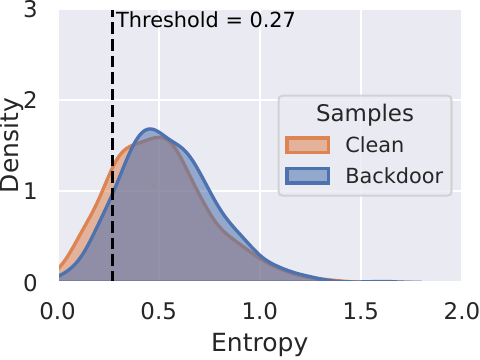}
    \caption{INS}
    \label{fig:strip_kde_ins2}
  \end{subfigure}
  \hfill
  \begin{subfigure}[htbp]{.155\textwidth}
    \centering
    \includegraphics[width=\linewidth]{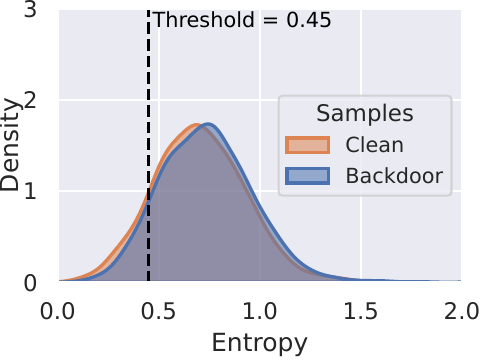}
    \caption{CON}
    \label{fig:strip_kde_con2}
  \end{subfigure}
  \caption{KDE plots of the entropy scores of more attacks under STRIP defense (complementary to~\Cref{fig:abl_strip_teco}).}
  \label{fig:strip_additional}
\end{figure}

\noindent \textbf{STRIP}~\cite{gaoSTRIPDefenceTrojan2019} monitors prediction consistency when noise is added to model inputs, using entropy as a measure of this consistency.
For clean inputs, perturbations produce random and varied predictions, yielding high entropy values.
In contrast, backdoored inputs exhibit low entropy due to their consistently biased predictions toward the target class.
To adapt STRIP for semantic segmentation models, we modify its approach from computing a single Shannon entropy value per input to generating an entropy map that captures per-pixel entropy across the segmentation mask.
We then aggregate these pixel-wise scores into a unified metric for the entire image.
This aggregated score serves as the decision criterion for identifying poisoned inputs.
We evaluate STRIP's detection performance using 250 clean and 250 poisoned samples, randomly sampled from the dataset.
The threshold is determined based on a target false positive rate (FPR): samples whose entropy lies below the FPR-th percentile of the clean entropy distribution are classified as poisoned samples.
\Cref{fig:strip_additional} presents additional KDE plots of STRIP entropies.
The results across different attacks indicate that the entropy distributions of clean and poisoned samples are highly overlapping and difficult to distinguish.
This suggests that entropy-based detection is ineffective for backdoor attacks in semantic segmentation.

\begin{table*}[t]
\scriptsize
  \centering
  \caption{Backdoor detection results under TeCo defense with additional evaluation metrics (complementary to~\Cref{tab:teco}).}
        \resizebox{0.65\linewidth}{!}{
    \begin{tabular}{c|cccc|cccc|cccc}
    \toprule
    \multicolumn{1}{c|}{\multirow{2}[4]{*}{\textbf{Attack}}} & \multicolumn{4}{c|}{\textbf{1std}} & \multicolumn{4}{c|}{\textbf{2std}} & \multicolumn{4}{c}{\textbf{3std}} \\
\cmidrule{2-13}          & \textbf{ACC} & \textbf{Recall} & \textbf{F1} & \textbf{AUC} & \textbf{ACC} & \textbf{Recall} & \textbf{F1} & \textbf{AUC} & \textbf{ACC} & \textbf{Recall} & \textbf{F1} & \textbf{AUC} \\
    \midrule
    HBA     & 0.568 & 0.2521 & 0.3494 & 0.5446 & 0.553 & 0.0652 & 0.1183 & 0.5169 & 0.540 & 0.0000     & 0.0000     & 0.5000 \\
    OFBA    & 0.583 & 0.2369 & 0.3433 & 0.5574 & 0.543 & 0.0630 & 0.1126 & 0.5074 & 0.540 & 0.0000     & 0.0000     & 0.5000 \\
    IBA     & 0.586 & 0.3112 & 0.4202 & 0.5765 & 0.529 & 0.0311 & 0.0599 & 0.5117 & 0.518 & 0.0000     & 0.0000     & 0.5000 \\
    O2O     & 0.528 & 0.1739 & 0.2532 & 0.5018 & 0.539 & 0.0478 & 0.0871 & 0.5026 & 0.540 & 0.0000     & 0.0000     & 0.5000 \\
    O2B     & 0.628 & 0.3826 & 0.4862 & 0.6098 & 0.560 & 0.0760 & 0.1373 & 0.5242 & 0.540 & 0.0000     & 0.0000     & 0.5000 \\
    B2O     & 0.558 & 0.1183 & 0.1754 & 0.4829 & 0.591 & 0.0377 & 0.0683 & 0.4965 & 0.603 & 0.0000     & 0.0000     & 0.5000 \\
    B2B     & 0.604 & 0.1445 & 0.1951 & 0.4885 & 0.655 & 0.0271 & 0.0496 & 0.4971 & 0.668 & 0.0000     & 0.0000     & 0.5000 \\
    INS     & 0.572 & 0.2134 & 0.3089 & 0.5312 & 0.548 & 0.0589 & 0.1067 & 0.5098 & 0.540 & 0.0000     & 0.0000     & 0.5000 \\
    CON     & 0.595 & 0.2687 & 0.3712 & 0.5523 & 0.556 & 0.0698 & 0.1245 & 0.5187 & 0.540 & 0.0000     & 0.0000     & 0.5000 \\

    \bottomrule
    \end{tabular}%
    }
  \label{tab:teco_additional}%
\end{table*}%

\begin{figure}[t]
  \scriptsize
  \begin{subfigure}[htbp]{.155\textwidth}
    \centering
    \includegraphics[width=\linewidth]{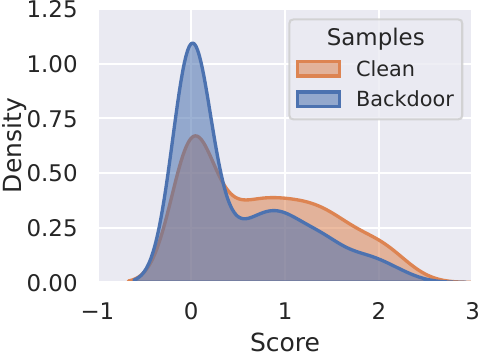}
    \caption{HBA}
    \label{fig:teco_kde_hba2}
  \end{subfigure}
  \hfill
  \begin{subfigure}[htbp]{.155\textwidth}
    \centering
    \includegraphics[width=\linewidth]{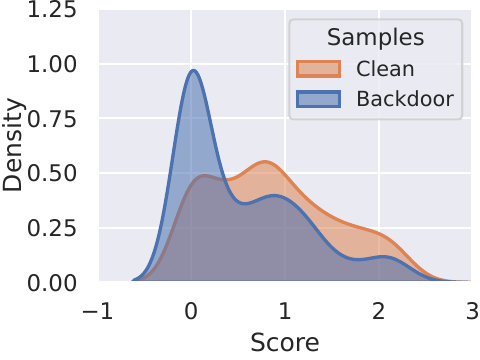}
    \caption{OFBA}
    \label{fig:teco_kde_ofba2}
  \end{subfigure}
  \hfill
  \begin{subfigure}[htbp]{.155\textwidth}
    \centering
    \includegraphics[width=\linewidth]{figures/fig_teco_kde_iba.pdf}
    \caption{IBA}
    \label{fig:teco_kde_iba2}
  \end{subfigure}
  \vfill
  \begin{subfigure}[htbp]{.155\textwidth}
    \centering
    \includegraphics[width=\linewidth]{figures/fig_teco_kde_o2o.pdf}
    \caption{O2O}
    \label{fig:teco_kde_o2o2}
  \end{subfigure}
  \hfill
  \begin{subfigure}[htbp]{.155\textwidth}
    \centering
    \includegraphics[width=\linewidth]{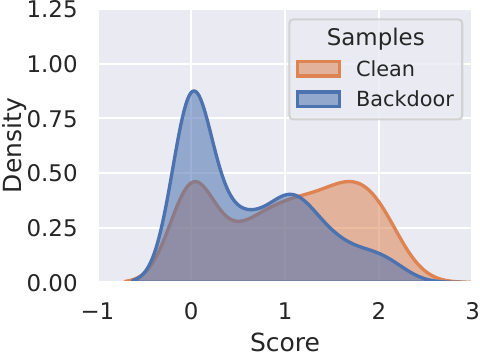}
    \caption{O2B}
    \label{fig:teco_kde_o2b2}
  \end{subfigure}
  \hfill
  \begin{subfigure}[htbp]{.155\textwidth}
    \centering
    \includegraphics[width=\linewidth]{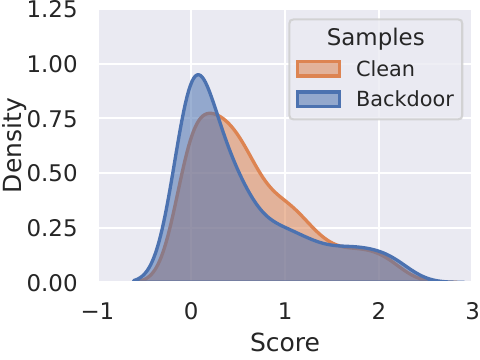}
    \caption{B2O}
    \label{fig:teco_kde_b2o2}
  \end{subfigure}
  \vfill
  \begin{subfigure}[htbp]{.155\textwidth}
    \centering
    \includegraphics[width=\linewidth]{figures/fig_teco_kde_b2b.pdf}
    \caption{B2B}
    \label{fig:teco_kde_b2b2}
  \end{subfigure}
  \hfill
  \begin{subfigure}[htbp]{.155\textwidth}
    \centering
    \includegraphics[width=\linewidth]{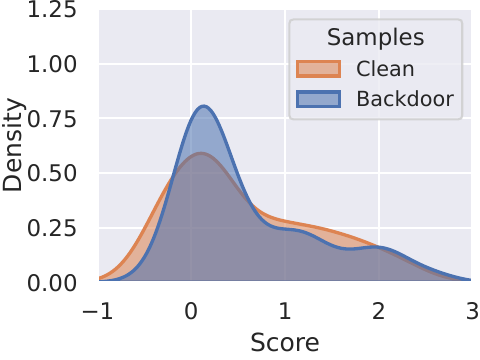}
    \caption{INS}
    \label{fig:teco_kde_ins2}
  \end{subfigure}
  \hfill
  \begin{subfigure}[htbp]{.155\textwidth}
    \centering
    \includegraphics[width=\linewidth]{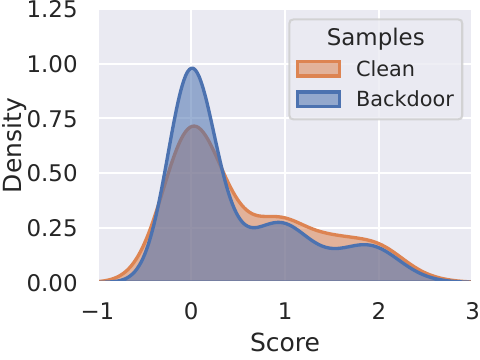}
    \caption{CON}
    \label{fig:teco_kde_con2}
  \end{subfigure}
  \caption{KDE plots of the detection scores of more attacks under TeCo defense (complementary to~\Cref{fig:abl_strip_teco}).}
  \label{fig:teco_additional}
\end{figure}

\noindent \textbf{TeCo}~\cite{liuDetectingBackdoorsInference2023} operates on the principle that clean images exhibit consistent robustness patterns across different types of image corruptions.
These corruptions include noise, blur, brightness changes, pixel-level perturbations, and image compression.
For instance, an image robust to noise corruption typically demonstrates predictable robustness to blur corruption as well.
Conversely, backdoor triggers exhibit inconsistent robustness profiles across different types of corruption.
They may be resilient to certain corruptions while being highly sensitive to others, resulting in high variance in the severity required to disrupt the model's prediction.
To adapt TeCo for semantic segmentation models, we need to consider whether all labels on the prediction mask have changed.
We compute the mIoU between the clean prediction mask and the corrupted prediction mask.
A prediction is considered ``broken'' when the mIoU falls below a predefined threshold, indicating substantial degradation in segmentation quality under corruption.
We adopt the 15 corruption types used in~\cite{liuDetectingBackdoorsInference2023} and report the corresponding results, including \emph{gaussian noise}, \emph{shot noise}, \emph{impulse noise}, \emph{defocus blur}, \emph{glass blur}, \emph{motion blur}, \emph{zoom blur}, \emph{snow}, \emph{frost}, \emph{fog}, \emph{brightness}, \emph{contrast}, \emph{elastic transform}, \emph{pixelate}, and \emph{jpeg compression}.
We evaluate TeCo's detection performance using 500 clean and 500 poisoned samples.
\Cref{tab:teco_additional} provides additional experimental results on defense methods.
\Cref{fig:teco_additional} presents additional KDE plots of TeCo scores.
The results across different attacks indicate that the score distributions of clean and poisoned samples are also highly overlapping.
This suggests that TeCo is ineffective for segmentation backdoor attacks.

\noindent \textbf{Beatrix}~\cite{maBeatrixResurrectionsRobust2023} analyzes subtle changes in a model's internal activation patterns using Gramian information.
The method operates on the principle that backdoor triggers induce statistically significant anomalies in the internal feature correlations of a model.
It exploits this discrepancy by modelling the Gramian information of feature maps to distinguish between clean and poisoned samples.
Adapting Beatrix to semantic segmentation models requires a class-wise sample grouping approach.
In classification models, Beatrix leverages ground truth class labels to group input samples.
It then calculates Gramian information within each group to identify poisoned samples.
However, semantic segmentation models produce masks containing per-pixel class labels.
This differs from single categorical predictions in classification tasks.
To accommodate this structural difference, we adopt a straightforward adaptation strategy.
For each ground truth segmentation mask, we identify the most dominant class, which is the class occupying the largest pixel area.
We then use this dominant class as the grouping criterion for that sample.
We evaluate Beatrix's detection performance on 100 clean and 100 poisoned samples.
The experimental results in~\Cref{sec:defenses} demonstrate that Beatrix is ineffective for segmentation backdoor attacks.

\section{Additional Discussion on Attacking Transformers}
\label{sec:transformers_additional}

Prior work has shown that Transformer-based models are vulnerable to a range of privacy and security attacks~\cite{zhangHowDoesDeep2024a,wangCanCNNsBe2023}.
\Cref{tab:exp_transformers} reports backdoor attack results for ViT-B~\cite{dosovitskiyImageWorth16x162021}, DeiT-S~\cite{touvronTrainingDataefficientImage2021}, and Swin-T~\cite{liuSwinTransformerHierarchical2021} on BDD100K.
Across all settings, these transformers are comparably vulnerable to backdoor attacks on conventional segmentation models.

\noindent \textbf{Performance of Swin-T.}
Swin-T and ConvNeXt-T both preserve strong local inductive biases and hierarchical multi-scale representations.
ConvNeXt-T extracts local patterns through convolutions, while the Swin Transformer relies on windowed self-attention.
In both cases, a localized trigger can be reliably captured in the early layers and passed through the feature pyramid to the segmentation head.
This locality and multi-scale feature extraction enable learning a stable trigger-to-mask manipulation, leading to comparable attack performance.

\noindent \textbf{Performance of ViT-B and DeiT.}
ViT-B and DeiT tokenize images into patches, emphasizing global token interactions that can weaken the signal of small, localized triggers.
The trigger may not consistently dominate attention across diverse scenes, making it harder for the model to learn a highly reliable backdoor.
Nevertheless, their ASRs remain non-trivial, indicating that they are still clearly susceptible to backdoor attacks.

\begin{figure*}[t]
  \scriptsize
  \centering
  \includegraphics[width=0.85\linewidth]{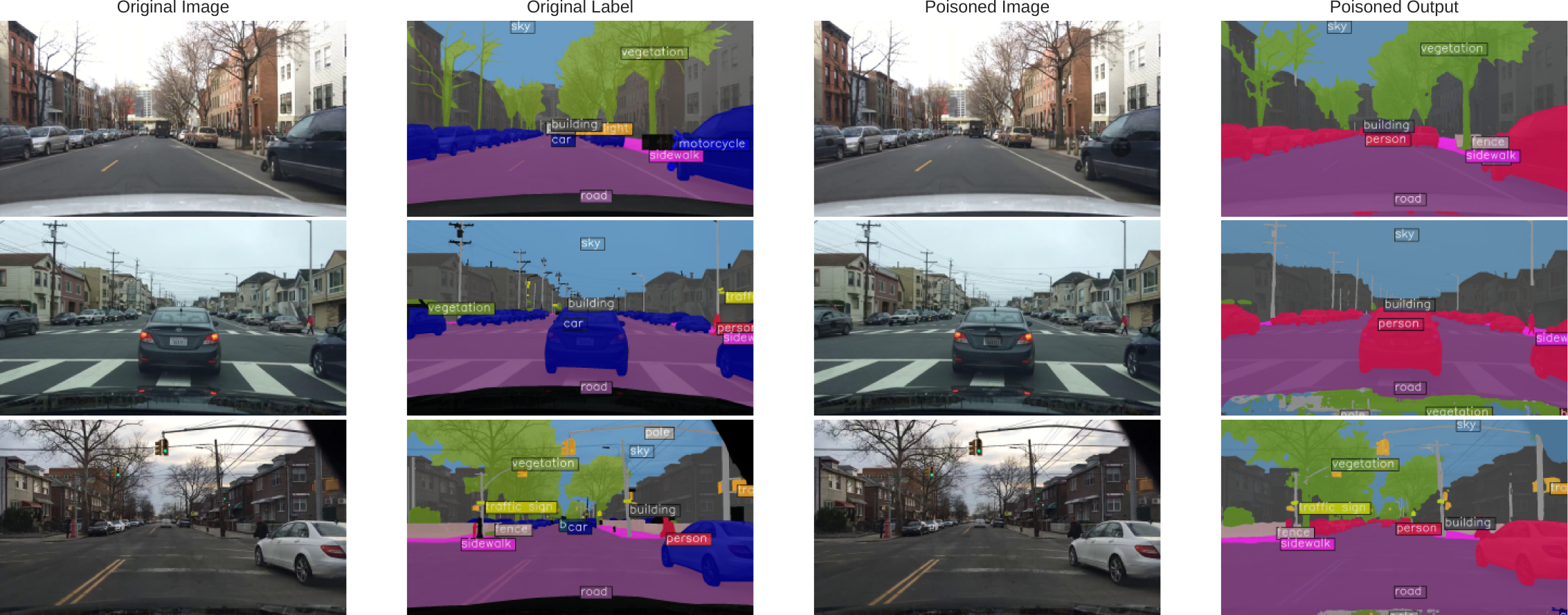}
  \caption{Visualization for O2O Attack.}
  \label{fig:vis_o2o}
\end{figure*}

\begin{figure*}[t]
  \scriptsize
  \centering
  \includegraphics[width=0.85\linewidth]{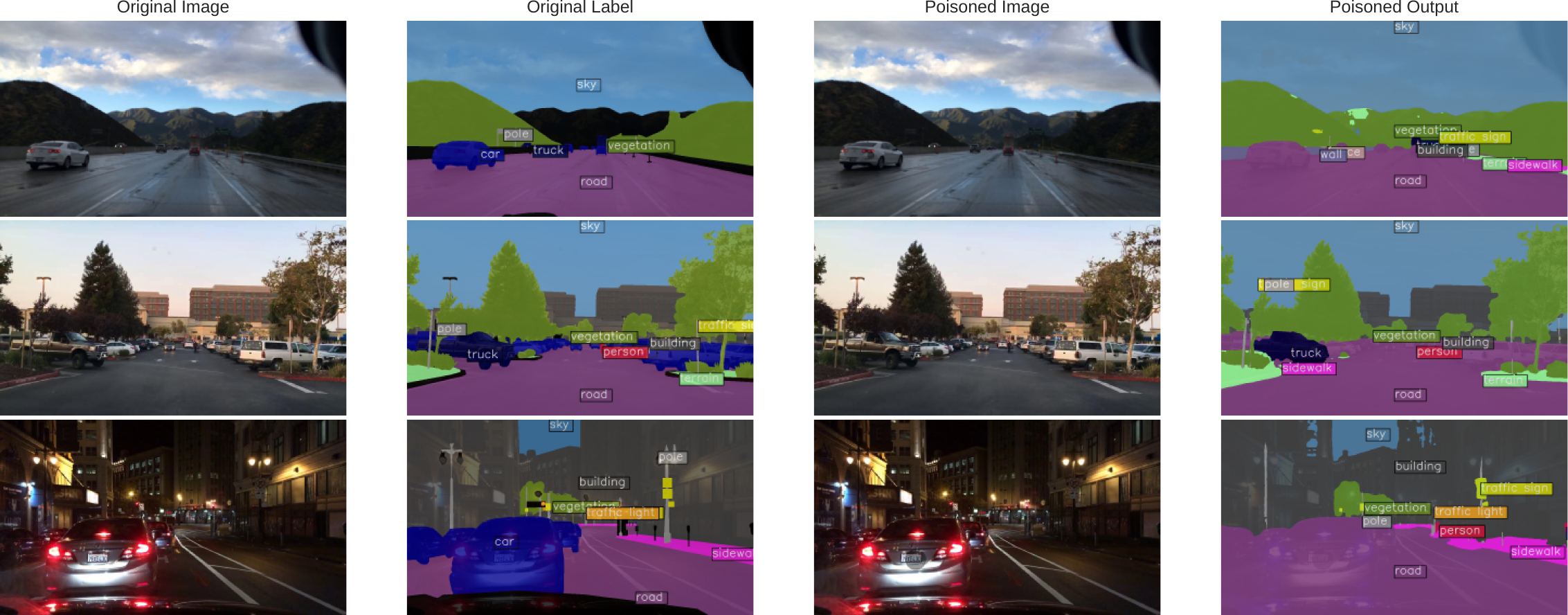}
  \caption{Visualization for O2B Attack.}
  \label{fig:vis_o2b}
\end{figure*}

\begin{figure*}[t]
  \scriptsize
  \centering
  \includegraphics[width=0.85\linewidth]{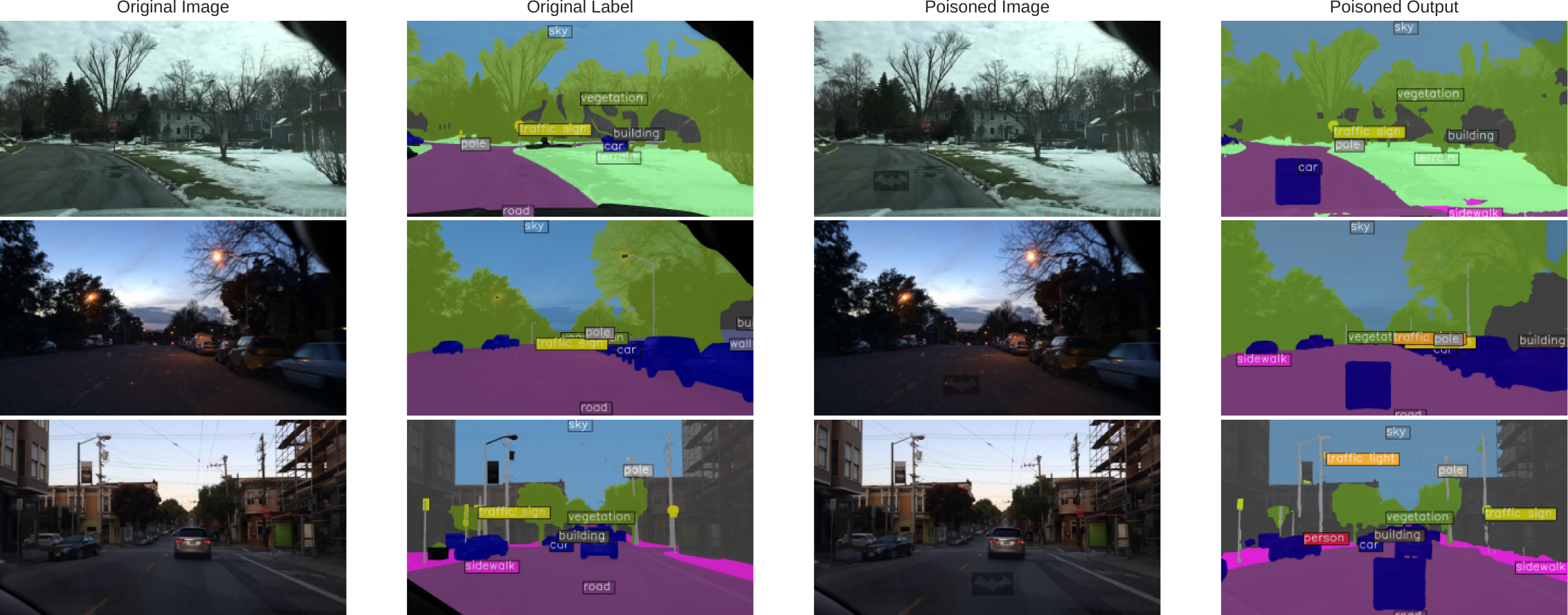}
  \caption{Visualization for B2O Attack.}
  \label{fig:vis_b2o}
\end{figure*}

\begin{figure*}[t]
  \scriptsize
  \centering
  \includegraphics[width=0.85\linewidth]{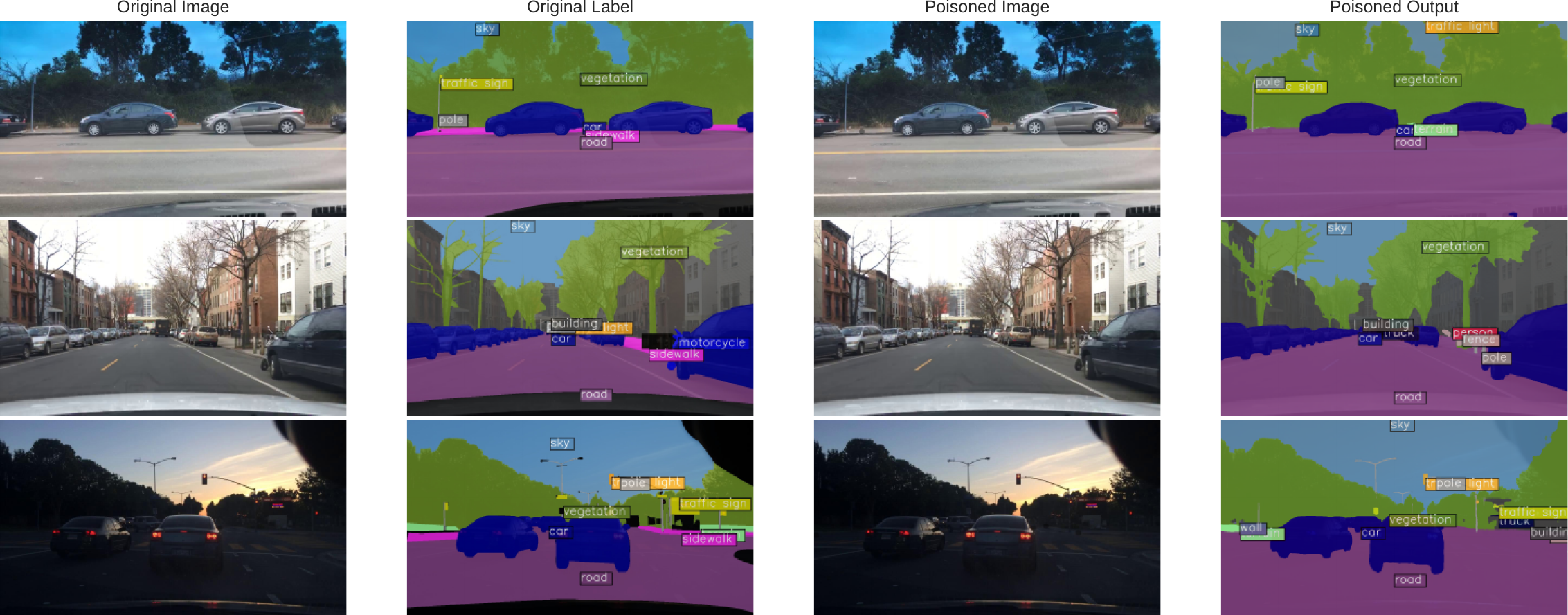}
  \caption{Visualization for B2B Attack.}
  \label{fig:vis_b2b}
\end{figure*}

\begin{figure*}[t]
  \scriptsize
  \centering
  \includegraphics[width=0.85\linewidth]{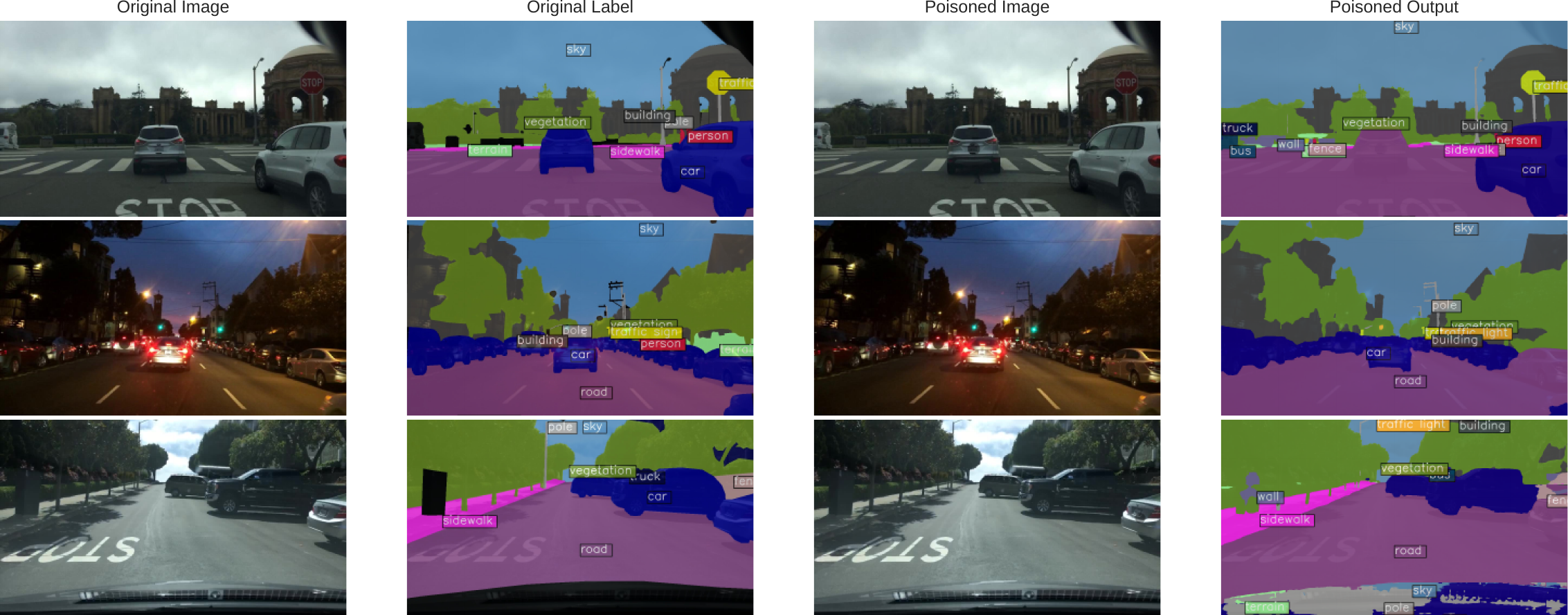}
  \caption{Visualization for Instance-Level Attack.}
  \label{fig:vis_ins}
\end{figure*}

\begin{figure*}[t]
  \scriptsize
  \centering
  \includegraphics[width=0.85\linewidth]{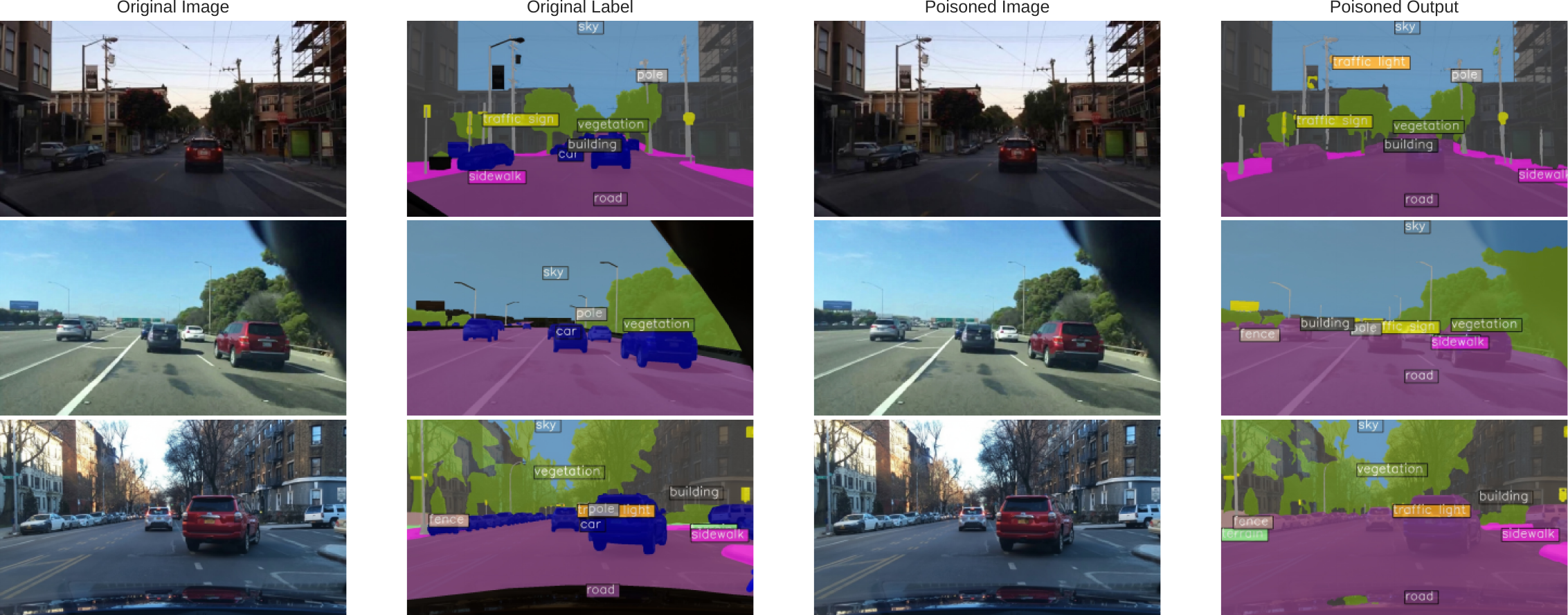}
  \caption{Visualization for Conditional Attack.}
  \label{fig:vis_con}
\end{figure*}

\section{Visualization}

\Cref{fig:vis_o2o,fig:vis_o2b,fig:vis_b2o,fig:vis_b2b,fig:vis_ins,fig:vis_con} visualize our attacks on BDD100K, with one representative setting for each attack vector.
In the O2O attack (\Cref{fig:vis_o2o}), the model maps the victim class \emph{car} to the target class \emph{person} once the trigger is present.
In the O2B attack (\Cref{fig:vis_o2b}), the model maps \emph{car} to the background class \emph{road}.
In the B2O attack (\Cref{fig:vis_b2o}), the model instead maps the background class \emph{road} to the object class \emph{car}, producing hallucinated object regions.
In the B2B attack (\Cref{fig:vis_b2b}), the model maps \emph{sidewalk} to \emph{road} when triggers are injected into the victim region.
In the INS attack (\Cref{fig:vis_ins}), the trigger affects only the stamped object instance: the attacked \emph{car} instance is flipped to \emph{road}, while other \emph{car} instances remain unchanged.
In the CON attack (\Cref{fig:vis_con}), the backdoor behavior is conditioned on context: the model maps \emph{car} to \emph{road} only when the trigger appears on a red car.
Overall, these visualizations confirm the effectiveness of our attacks and illustrate how triggers can induce targeted yet context-dependent segmentation errors.

\section{Additional Related Work}
\label{sec:related_work_additional}

\noindent \textbf{Other Related Research.}
Some research explored backdoor triggers embedded within objects in images~\cite{chanBadDetBackdoorAttacks2022,chenCleanimageBackdoorAttacking2023,chengOdScanBackdoorScanning2024,luAnywhereDoorMultiTargetBackdoor2024}, which relates to our coarse-grained attacks.
These works primarily target object detection by manipulating bounding box predictions.
In contrast, our attacks focus on semantic segmentation, manipulating individual pixel classifications to produce incorrect segmentation masks, requiring distinct trigger design strategies.

Some prior backdoor attacks~\cite{duanConditionalBackdoorAttack2024,shafahiPoisonFrogsTargeted2018} share similarities with our fine-grained attacks, but key distinctions remain.
In~\cite{duanConditionalBackdoorAttack2024}, they use a single condition (JPEG compression presence).
However, our method requires composite conditions, which involve both a specific trigger pattern on an object and the object having particular attributes (i.e., a specific color).
This enables more flexible, context-dependent attacks with enhanced stealthiness.
In~\cite{shafahiPoisonFrogsTargeted2018}, they focused on model degradation rather than backdoor activation, whereas our method targets backdoor attacks through trigger and object associations.

\section{Discussion on Real-World Implementation}

Our work identifies four coarse-grained attack vectors, defined by class-level semantic labels, and two fine-grained vectors, characterised by activation specificity.
We discuss how these attacks could activate in a realistic autonomous-driving perception pipeline, where a backdoored segmentation model processes images captured by a front-facing camera.

We consider a practical adversary who can physically place a small trigger in the environment (e.g., near a roadway) but does not need to tamper with the victim vehicle.
Once the trigger enters the camera's field of view, the backdoored model may produce a manipulated segmentation mask while behaving normally on clean scenes.
In practice, the geometric trigger can be printed (e.g., on paper or a sticker) and attached to common surfaces or objects with minimal effort.

For O2O, O2B and INS attacks, the trigger can be placed on the rear of a vehicle parked by the roadside, such that it appears in the captured scene when the victim car passes by.
For B2O and B2B attacks, the trigger can be placed on planar background regions such as the road surface or sidewalk to induce false positives or background relabeling.
For the CON attack, the trigger can be attached to a specific contextual object (e.g., the rear of a red car), causing activation only when the contextual condition is satisfied.
As the victim vehicle approaches and captures the scene, the trigger becomes visible in one or more frames, activating the backdoor and causing the attacker-chosen manipulation.

\end{document}